\documentclass[
 reprint,
superscriptaddress,
 amsmath,amssymb,
 aps, nofootinbib
]{revtex4-2}

\usepackage{graphicx}
\usepackage{dcolumn}
\usepackage{bm}
\usepackage{float}
\usepackage{epsfig}
\usepackage{amsfonts}
\usepackage{amssymb}
\usepackage{amsmath}
\usepackage{mathtools}
\usepackage{multirow}
\usepackage{lipsum}
\usepackage{physics}

\usepackage{rsfso}
\usepackage{mathrsfs}
\usepackage{booktabs}
\usepackage[normalem]{ulem}
\usepackage{placeins}
\usepackage{xcolor}
\usepackage{MnSymbol,wasysym}
\DeclareMathAlphabet{\mathcal}{OMS}{cmsy}{m}{n}
\usepackage{hyperref}
\hypersetup{colorlinks=false,breaklinks=true,hidelinks}

\newcommand{\V}{\mathcal{V}}
\usepackage{cancel}

\newcommand{\bk}{\boldsymbol{k}}
\newcommand{\bp}{\boldsymbol{p}}

\newcommand{\bx}{\boldsymbol{x}}
\newcommand{\by}{\boldsymbol{y}}

\newcommand{\de}{{\rm d}}
\def\bea{\begin{eqnarray}}
\def\eea{\end{eqnarray}}
\def\be{\begin{equation}}
\def\ee{\end{equation}}
\def\ba{\begin{array}}
\def\ea{\end{array}}
\def\nn{\nonumber}

\usepackage{tikz}
\usetikzlibrary{calc}
\usetikzlibrary{arrows,patterns}

\begin{document}

\title{Regularizing infrared divergences in de Sitter spacetime: \\ Loops, dimensional regularization, and cutoffs}

\author{Javier Huenupi}
 \email{javier.huenupi@ug.uchile.cl}
  \affiliation{Departamento de F\'isica, Facultad de Ciencias F\'isicas y Matem\'aticas, Universidad de Chile, Santiago, Chile}
\author{Ellie Hughes}
 \email{ellieh@mit.edu}
  \affiliation{Center for Theoretical Physics, Massachusetts Institute of Technology, Cambridge, Massachusetts 02139, USA}
  \affiliation{Departamento de F\'isica, Facultad de Ciencias F\'isicas y Matem\'aticas, Universidad de Chile, Santiago, Chile}
\author{Gonzalo A. Palma}
 \email{gpalmaquilod@ing.uchile.cl}
  \affiliation{Departamento de F\'isica, Facultad de Ciencias F\'isicas y Matem\'aticas, Universidad de Chile, Santiago, Chile}
\author{Spyros Sypsas}
 \email{s.sypsas@gmail.com}
  \affiliation{High Energy Physics Research Unit, Faculty of Science, Chulalongkorn University, Bangkok 10330, Thailand}
  \affiliation{National Astronomical Research Institute of Thailand, Don Kaeo, Mae Rim, Chiang Mai 50180, Thailand}

\date{January 3, 2025}

\begin{abstract}

{Correlation functions of light scalar fields in de Sitter spacetime, computed via standard perturbation theory, often exhibit secular growth characterized by time-dependent divergent terms in the form of powers of $\ln a(t)$, where $a(t)$ is the scale factor describing cosmic expansion. It is widely believed that loop corrections further enhance this secular growth. We argue that this is not necessarily the case: Loop corrections can be systematically handled using standard perturbative techniques, such as dimensional regularization, without introducing new $\ln a(t)$ terms. We focus on a canonical massless scalar field $\varphi$ with self-interactions described by a potential $\mathcal{V}(\varphi)$, and analyze correlation functions represented by diagrams with a single vertex and an arbitrary number of loops. In this framework, infrared divergences can be systematically eliminated with counterterms at each order in perturbation theory, leading to loop-corrected correlation functions that are indistinguishable from their tree-level forms, with no secular growth from loops. Furthermore, adopting a Wilsonian perspective, we explore the role of cutoffs in computing loop corrections within effective field theory and identify the effective potential $\mathcal{V}_{\rm eff}(\varphi)$, which guarantees cutoff-independent observables. We conclude that when infrared comoving cutoffs are used to regularize loop integrals, time-dependent Wilsonian coefficients are necessary to maintain cutoff-free correlation functions. Neglecting this time dependence results in secular growth from loops.

}

\end{abstract}

\maketitle

\section{Introduction} \label{sec:introduction}

The aim of this paper is to comment on the status of loop corrections to $n$-point correlation functions of light scalar fields in de Sitter space~\cite{Vilenkin:1982wt,Linde:1982uu,Starobinsky:1982ee,Vilenkin:1983xp,Tsamis:2005hd,Tsamis:1993ub,Brunier:2004sb,Onemli:2002hr,Miao:2005am,Sloth:2006az,Sloth:2006nu,Riotto:2008mv,Seery:2007we,Bartolo:2007ti,Finelli:2008zg,Rajaraman:2010zx,Seery:2010kh,Xue:2011hm,Akhmedov:2013vka,Akhmedov:2013xka,Akhmedov:2017ooy,Akhmedov:2019cfd,Cespedes:2023aal,Baumgart:2019clc,Sasaki:1992ux,Bilandzic:2007nb,Enqvist:2008kt,Weinberg:2005vy,Weinberg:2006ac,Adshead:2008gk,Chen:2018brw,Chen:2018uul,Palma:2023idj,Palma:2023uwo,Burgess:2009bs,Burgess:2010dd,Negro:2024bbf,Urakawa:2009gb,Urakawa:2009my,Tanaka:2013caa,Senatore:2009cf,Gorbenko:2019rza}. Non-derivative interactions---those descending from scalar field potentials---lead to the appearance of secular growth in light-field $n$-point correlation functions in momentum space~\cite{Vilenkin:1982wt,Linde:1982uu,Starobinsky:1982ee,Vilenkin:1983xp}. This growth takes the form of infrared asymptotically diverging terms consisting in powers of $\ln a(t)$, where $a(t)$ is the scale factor parametrizing the expansion of flat spatial slices. Already at tree level, $n$-point correlation functions in momentum space turn out to be proportional to $[ \ln a(t) ]^V$, with $V$ the number of vertices participating in a contributing diagram~\cite{Baumgart:2019clc, Palma:2023idj}. Do loop corrections alter this behavior? The answer to this question is often found to be positive~\cite{Vilenkin:1982wt,Linde:1982uu,Starobinsky:1982ee,Vilenkin:1983xp,Tsamis:2005hd,Tsamis:1993ub,Brunier:2004sb,Onemli:2002hr,Miao:2005am,Sloth:2006az,Sloth:2006nu,Riotto:2008mv,Seery:2007we,Bartolo:2007ti,Rajaraman:2010zx,Seery:2010kh,Finelli:2008zg,Baumgart:2019clc,Cespedes:2023aal,Sasaki:1992ux,Bilandzic:2007nb,Enqvist:2008kt,Weinberg:2005vy,Weinberg:2006ac,Adshead:2008gk,Xue:2011hm,Akhmedov:2013vka,Akhmedov:2013xka,Akhmedov:2017ooy,Akhmedov:2019cfd}, resulting from the way in which loop integrals are dealt with. We wish to revisit this argument.

It is well known that massless scalar fields do not allow the construction of normalizable de Sitter-invariant vacuum states~\cite{Allen:1985ux,Allen:1987tz}. This issue arises from the lack of a gap between the vacuum state and zero-momentum multi-particle states, which, in turn, causes divergences in propagators (written in coordinate space) due to unbounded zero-mode contributions. Let us recall that propagators participate in the computation of correlation functions in two distinct ways: as bulk-to-boundary propagators and as bulk-to-bulk propagators. In the former class, infrared divergences are not a concern. This is simply because the range of wavelengths to which we have observational access is finite, which, in particular, excludes zero modes. In practice, correlation functions participating in the computation of observables must be filtered via window functions restricting the range of external momenta flowing into the contributing diagrams. Depending on the case under consideration, this filtering can be implemented in different ways. A typical choice consists of screening bulk-to-boundary propagators from infinitely long modes with a window function featuring a comoving infrared momentum cutoff. 

On the contrary, infrared divergences emerging from bulk-to-bulk propagators are a real concern, as they lead to the appearance of unbounded loop integrals exhibiting both infrared (IR) and ultraviolet (UV) divergences. A common regularization method is the introduction of cutoffs. The main purpose of this article is to highlight the nature and implications of the infrared momentum cutoff used to regularize these loop corrections. 

By far, the most common approach found in the literature to regularize loop integrals consists in the use of an IR cutoff scale set to be a constant in comoving coordinates~\cite{Vilenkin:1982wt,Linde:1982uu,Starobinsky:1982ee,Vilenkin:1983xp,Tsamis:2005hd,Tsamis:1993ub,Brunier:2004sb,Onemli:2002hr,Miao:2005am,Sloth:2006az,Sloth:2006nu,Riotto:2008mv,Seery:2007we,Bartolo:2007ti,Rajaraman:2010zx,Seery:2010kh,Xue:2011hm,Akhmedov:2013vka,Akhmedov:2013xka,Akhmedov:2017ooy,Akhmedov:2019cfd,Finelli:2008zg,Baumgart:2019clc,Cespedes:2023aal,Sasaki:1992ux,Bilandzic:2007nb,Enqvist:2008kt,Weinberg:2005vy,Weinberg:2006ac,Adshead:2008gk} (recall that physical momentum $\bp$ is related to comoving momentum $\bk$ via ${\bp} = {\bk} / a(t)$. Hence, a constant comoving cutoff implies a time-dependent physical cutoff and vice versa). The propagator resulting from this choice becomes finite but is found to contribute extra time dependencies to correlators~\cite{Onemli:2002hr}, now scaling as $[\ln a(t)]^{V+L}$. Here, $L$ represents the loop number in a diagram with $V$ vertices. Conversely, if the infrared cutoff is set to be constant in physical coordinates~\cite{Burgess:2009bs,Burgess:2010dd,Negro:2024bbf,Chen:2018brw,Chen:2018uul,Palma:2023idj,Palma:2023uwo}, the internal propagators remain finite while correlators maintain their tree-level temporal behavior of $[\ln a(t)]^V$. 

It is worth noting a prevailing rationale favoring the adoption of comoving cutoffs: the notion that inflation must possess an origin~\cite{Vilenkin:1982wt,Linde:1982uu,Starobinsky:1982ee,Vilenkin:1983xp}, and, therefore, a comoving boundary signaling the physical size of the expanding universe at some initial time must be taken into account in the computation of correlators. While this statement might have merit, we observe that, sooner or later, this comoving boundary will reach the infrared physical length scale determining the validity of perturbation theory. As such, the details of how the universe looks on scales much larger than this length scale cannot be relevant for the computation of correlation functions using perturbation theory towards the infrared. 

Here, we argue that loops, properly treated, cannot increase the infrared divergence of diagrams. Our argument is simple: infrared divergences arising from loop integrals can be regularized with the help of dimensional regularization without resorting to cutoffs. Hence, any method introducing cutoffs must agree with this cutoff free approach. As we shall see, at first sight this result favors the use of constant cutoffs in physical momentum space (in accordance with Refs.~\cite{Burgess:2009bs,Burgess:2010dd,Negro:2024bbf}). However, we will conclude that the choice between physical and comoving cutoffs is in fact irrelevant. That is, loop corrections computed with comoving cutoffs---properly treated---cannot lead to results differing from the cutoff-free approach offered by dimensional regularization, and thus comoving cutoffs cannot enhance secular growth. We will examine these statements in detail for the case of diagrams with one vertex and an arbitrary number of loops, an example that is particularly simple to solve exactly because these loops carry no external momenta. Even though this case might seem too limited to draw useful general conclusions, we notice that in this example we already obtain results that depart from the standard lore where loops contribute to the enhancement of infrared divergences. Moreover, the key elements allowing us to appreciate that these statements should hold true in more general diagrams are already present in the analysis of single-vertex diagrams.

In a nutshell, we will offer a three-pronged analysis to show that infrared divergences appearing in propagators can be safely treated with standard perturbative techniques. First, we show that infrared divergences due to bulk-to-bulk propagators appearing in loops can be absorbed via counterterms within standard schemes such as dimensional regularization. The resulting correlation functions turn out to be finite and independent of any choice of cutoffs (comoving or physical). In the second approach, we adopt a Wilsonian perspective and define an effective field theory (EFT) with infrared and ultraviolet cutoff scales defined in physical momentum space. By requiring that observable correlation functions must remain independent of these scales, we recover the same conditions encountered in the first analysis based on dimensional regularization. That is,  by choosing physical cutoffs, the resulting effective theory retains the critical property of time-translation invariance---a manifest property of de Sitter spacetime---already present in the starting bare theory. Last but not least, we consider what happens with the EFT employed to compute correlation functions if one chooses to work with a constant IR cutoff in comoving space (and keep the UV cutoff to be a constant in physical space). We find that Wilsonian coefficients must necessarily be time dependent in order to keep observables independent of cutoffs. This result clarifies why the introduction of comoving IR cutoffs has consistently led to the appearance of additional secular growth: time-dependent Wilsonian coefficients were omitted.

To conclude this preamble, let us emphasize that the statement that loops in de Sitter correlators regulated with constant comoving IR cutoffs must be considered with caution is not new. (See, for instance, Refs.~\cite{Burgess:2009bs,Burgess:2010dd,Negro:2024bbf,Urakawa:2009gb,Urakawa:2009my,Tanaka:2013caa} for different reasons why such a cutoff may lead to spurious results.) Furthermore, as has been argued in Ref.~\cite{Senatore:2009cf},  the distinction between comoving and physical UV cutoffs significantly impacts the computation of observables. Given the difficulty of determining the correct form of time-dependent Wilsonian coefficients, in this article we aim to provide new arguments in favor of using constant physical IR cutoffs to regulate loops or, alternatively, employing dimensional regularization, which eliminates the need for cutoffs.\footnote{Another way to compute correlators is via the wavefunction of the universe approach~\cite{Maldacena:2002vr,Anninos:2014lwa,Pajer:2020wxk,Arkani-Hamed:2018kmz}. The result of Ref.~\cite{Anninos:2014lwa} for the 1-loop, 2-point function agrees with the result of Refs.~\cite{Burgess:2009bs,Burgess:2010dd}, which advocate a physical IR cutoff regulating the loop.}

\section{Correlation functions} \label{sec:correlators}

We will employ units such that $c=\hbar = 1$ and work in a de Sitter background spacetime described by a line element given by $\dd{s}^2=-\dd{t}^2+a^2(t)\dd{ \bx}^2$, where $a(t)\propto e^{Ht}$ is the scale factor and $H$ is the Hubble scale setting the de Sitter radius. We will also work with conformal time, hereby defined as $\tau = - 1 / H a(t)$, allowing one to rewrite the metric as $\dd{s}^2=a^2(\tau) \left( -\dd{\tau}^2+\dd{ \bx}^2 \right)$. We shall consider a standard canonical single field theory with a bare action of the form:
\begin{equation} \label{action-1}
   \!\! S\!=\!\int \!\!\dd[3]{x}\dd{\tau} a^4\left[\frac{\left(\varphi'\right)^2}{2a^2}-\frac{(\nabla\varphi)^2}{2a^2} - \frac{m^2\varphi^2}{2} -   \V_{\rm b}(\varphi)    \right],
\end{equation}
where $\V_{\rm b}(\varphi) $ represents an arbitrary bare potential and a prime ($'$) denotes the derivative with respect to $\tau$. To perform perturbation theory, we shall treat the entire potential $\V_{\rm b}(\varphi) $ as the expansion parameter in our perturbative scheme. (This statement will be made more precise once we define the renormalized potential $\V_{\rm ren}(\varphi)$ in Sec.~\ref{sec:dim-reg}.)  In addition to $\V_{\rm b}(\varphi) $, we consider a mass parameter $m$ that will help us to tackle separately the distinct situations where the field is massive and exactly massless. We take the term $\frac{1}{2} m^2 \varphi^2$ as part of the free theory, whereas the quadratic part of the potential $\frac{1}{2} \lambda_{2}^{\rm b} \varphi^2 \subset \mathcal V_{\rm b} (\varphi)$ [see Eq.~\eqref{eqn:Vb}] is taken to be a small perturbative parameter leading to the definition of two-legged vertices entering the computation of correlation functions.

The general quantized solution for $\varphi ({\bx} , \tau)$ may be written as
\be
\varphi ({\bx} , \tau) = \int \frac{\dd[3] k}{(2\pi)^3} e^{- i {\bk} \cdot {\bx}} \varphi_{\bk} (\tau) ,
\ee
where $\varphi_{\bk} (\tau)$ is the field in comoving momentum space. At zeroth order ($\mathcal V_{\rm b} (\varphi) = 0$) this quantity may be written as $\varphi_{\bk} (\tau) =  f_k(\tau) \hat a_{\bk}  +  f^*_k(\tau) \hat a^{\dag}_{-\bk}$, where $\hat a^{\dag}_{\bk}$ and $\hat a_{\bk}$ are creation and annihilation operators satisfying $\big[\hat a_{\bk} , \hat a^{\dag}_{\bk'} \big] = (2 \pi)^3 \delta^{(3)} ({\bk} - {\bk}')$, and $f_k (\tau)$ represents a mode function satisfying the free-field equations of motion in comoving momentum space, with Bunch-Davies initial conditions. In the case of a massless scalar field, it is given by
\begin{equation} 
    f_k(\tau) =  \frac{H}{ \sqrt{2 k^3}}  ( 1 + i k \tau) e^{- i k \tau},
\end{equation}
whereas, in the more general case of a massive field, it is given by
\begin{equation} 
    f_k(\tau) = \frac{i\sqrt{\pi}H}{2 k^{3/2}} (-k \tau )^{3/2} H_\nu^{(1)}(-k\tau),
\end{equation}
where $H_\nu^{(1)}(x)$ is a Hankel function of the first kind with $\nu = \sqrt{9/4 - m^2/H^2}$.

In order to compute correlation functions perturbatively, we expand the bare potential $\V_{\rm b}(\varphi)$ in a Taylor series:
\begin{equation}
    \V_{\rm b}(\varphi)=\sum_{n=0}^\infty\frac{\lambda_n^{\rm b}}{n!}\varphi^n ,
    \label{eqn:Vb}
\end{equation}
where $\lambda_n^{\rm b}$ denote arbitrary bare coupling constants. We may then resort to one of the many procedures available, such as the in-in  Hamiltonian scheme~\cite{Maldacena:2002vr,Weinberg:2005vy,Adshead:2009cb}, the in-in Schwinger-Keldysh approach~\cite{Calzetta:1986ey,Chen:2017ryl}, or the wavefunction of the universe formalism~\cite{Maldacena:2002vr,Anninos:2014lwa,Pajer:2020wxk,Arkani-Hamed:2018kmz}. In all of these approaches, the coupling $\lambda_n^{\rm b}$ defines an $n$-legged vertex allowing the building of diagrams. More to the point, to order $\lambda$, an $n$-point correlation function in comoving momentum space, evaluated at a time $\tau$, can be expressed as the following sum of single-vertex diagrams:
\usetikzlibrary {patterns.meta}
\tikzdeclarepattern{
  name=mylines,
  parameters={
      \pgfkeysvalueof{/pgf/pattern keys/size},
      \pgfkeysvalueof{/pgf/pattern keys/angle},
      \pgfkeysvalueof{/pgf/pattern keys/line width},
  },
  bounding box={
    (0,-0.5*\pgfkeysvalueof{/pgf/pattern keys/line width}) and
    (\pgfkeysvalueof{/pgf/pattern keys/size},
0.5*\pgfkeysvalueof{/pgf/pattern keys/line width})},
  tile size={(\pgfkeysvalueof{/pgf/pattern keys/size},
\pgfkeysvalueof{/pgf/pattern keys/size})},
  tile transformation={rotate=\pgfkeysvalueof{/pgf/pattern keys/angle}},
  defaults={
    size/.initial=5pt,
    angle/.initial=45,
    line width/.initial=.4pt,
  },
  code={
      \draw [line width=\pgfkeysvalueof{/pgf/pattern keys/line width}]
        (0,0) -- (\pgfkeysvalueof{/pgf/pattern keys/size},0);
  },
}
\def\treediagram{\tikz[baseline=-1.4ex]{
\coordinate (P) at (0,-3ex);
\coordinate (C) at (-9ex,4ex);
\draw (C) circle (0.01ex) node[anchor=east]{\footnotesize{$\tau$}};
\draw[thick,dashed] (-9ex,4ex) -- (9ex,4ex);
\draw[thick] (0,-3ex) -- (-8ex,4ex);
\draw[thick] (0,-3ex) -- (-4ex,4ex);
\draw[thick] (0,-3ex) -- (4ex,4ex);
\draw[thick] (0,-3ex) -- (8ex,4ex);
\filldraw[color=black, fill=black, thick] (-1.2ex,2ex) circle (0.1ex);
\filldraw[color=black, fill=black, thick] (0ex,2ex) circle (0.1ex);
\filldraw[color=black, fill=black, thick] (1.2ex,2ex) circle (0.1ex);
\node at ($(P) + (-0.5ex,0)$) [anchor=east]{\footnotesize{$\lambda^{\text{b}}_{n},\:\bar{\tau}$}};
\filldraw[color=black, fill=white, thick] (-8.5ex,3.5ex) rectangle (-7.5ex,4.5ex) node[anchor=south]{\footnotesize{$k_1$}};
\filldraw[color=black, fill=white, thick] (-4.5ex,3.5ex) rectangle (-3.5ex,4.5ex) node[anchor=south]{\footnotesize{$k_2$}};
\filldraw[color=black, fill=white, thick] (3.5ex,3.5ex) rectangle (4.5ex,4.5ex) node[anchor=south]{\footnotesize{$k_{n-1}$}};
\filldraw[color=black, fill=white, thick] (7.5ex,3.5ex) rectangle (8.5ex,4.5ex) node[anchor=south]{\footnotesize{$k_n$}};
\filldraw[color=white, fill=white] (P) circle (0.8ex);
\filldraw[pattern=north east lines, thick] (P) circle (0.8ex);
}
}
\def\oneloopdiagram{\tikz[baseline=-1.4ex]{
\coordinate (P) at (0,-3ex);
\coordinate (C) at (-9ex,4ex);
\draw (C) circle (0.01ex) node[anchor=east]{\footnotesize{$\tau$}};
\draw[thick,dashed] (-9ex,4ex) -- (9ex,4ex);
\draw[thick] (0,-3ex) -- (-8ex,4ex);
\draw[thick] (0,-3ex) -- (-4ex,4ex);
\draw[thick] (0,-3ex) -- (4ex,4ex);
\draw[thick] (0,-3ex) -- (8ex,4ex);
\filldraw[color=black, fill=black, thick] (-1.2ex,2ex) circle (0.1ex);
\filldraw[color=black, fill=black, thick] (0ex,2ex) circle (0.1ex);
\filldraw[color=black, fill=black, thick] (1.2ex,2ex) circle (0.1ex);
\node at ($(P) + (-0.5ex,0)$) [anchor=east]{\footnotesize{$\lambda^{\text{b}}_{n+2},\:\bar\tau$}};
\filldraw[color=black, fill=white, thick] (-8.5ex,3.5ex) rectangle (-7.5ex,4.5ex) node[anchor=south]{\footnotesize{$k_1$}};
\filldraw[color=black, fill=white, thick] (-4.5ex,3.5ex) rectangle (-3.5ex,4.5ex) node[anchor=south]{\footnotesize{$k_2$}};
\filldraw[color=black, fill=white, thick] (3.5ex,3.5ex) rectangle (4.5ex,4.5ex) node[anchor=south]{\footnotesize{$k_{n-1}$}};
\filldraw[color=black, fill=white, thick] (7.5ex,3.5ex) rectangle (8.5ex,4.5ex) node[anchor=south]{\footnotesize{$k_n$}};
\draw[thick,scale=3] (0,-1ex)  to[in=-70,out=-110,loop] (0,-1ex);
\filldraw[color=white, fill=white] (P) circle (0.8ex);
\filldraw[pattern=north east lines, thick] (P) circle (0.8ex);
}
}
\def\twoloopdiagram{\tikz[baseline=-1.4ex]{
\coordinate (P) at (0,-3ex);
\coordinate (C) at (-9ex,4ex);
\draw (C) circle (0.01ex) node[anchor=east]{\footnotesize{$\tau$}};
\draw[thick,dashed] (-9ex,4ex) -- (9ex,4ex);
\draw[thick] (0,-3ex) -- (-8ex,4ex);
\draw[thick] (0,-3ex) -- (-4ex,4ex);
\draw[thick] (0,-3ex) -- (4ex,4ex);
\draw[thick] (0,-3ex) -- (8ex,4ex);
\filldraw[color=black, fill=black, thick] (-1.2ex,2ex) circle (0.1ex);
\filldraw[color=black, fill=black, thick] (0ex,2ex) circle (0.1ex);
\filldraw[color=black, fill=black, thick] (1.2ex,2ex) circle (0.1ex);
\node at ($(P) + (-0.5ex,0)$) [anchor=east]{\footnotesize{$\lambda^{\text{b}}_{n+4},\:\bar{\tau}$}};
\filldraw[color=black, fill=white, thick] (-8.5ex,3.5ex) rectangle (-7.5ex,4.5ex) node[anchor=south]{\footnotesize{$k_1$}};
\filldraw[color=black, fill=white, thick] (-4.5ex,3.5ex) rectangle (-3.5ex,4.5ex) node[anchor=south]{\footnotesize{$k_2$}};
\filldraw[color=black, fill=white, thick] (3.5ex,3.5ex) rectangle (4.5ex,4.5ex) node[anchor=south]{\footnotesize{$k_{n-1}$}};
\filldraw[color=black, fill=white, thick] (7.5ex,3.5ex) rectangle (8.5ex,4.5ex) node[anchor=south]{\footnotesize{$k_n$}};
\draw[thick,scale=3] (0,-1ex)  to[in=-90,out=-130,loop] (0,-1ex);
\draw[thick,scale=3] (0,-1ex)  to[in=-50,out=-90,loop] (0,-1ex);
\filldraw[color=white, fill=white] (P) circle (0.8ex);
\filldraw[pattern=north east lines, thick] (P) circle (0.8ex);
}
}
\begin{widetext}
\begin{equation} \label{full-diagram}
\langle \varphi_{{\bk}_1} \cdots \,\varphi_{{\bk}_n} \rangle (\tau) = \treediagram + \oneloopdiagram + \twoloopdiagram + \cdots \;.
\end{equation}
\end{widetext}
Note that in the previous expression, an arbitrary diagram with $n$ external legs and $L$ loops must be proportional to $\lambda^{\rm b}_{n + 2L}$~\cite{Chen:2018brw}. Using, for instance, Feynman rules within the Schwinger-Keldysh formalism (see Appendix~\ref{app:feynman}), a single diagram with $L$ loops is found to have the following form:
\def\Oconnectedloopstwo{\tikz[baseline=-1.4ex]{
\coordinate (P) at (0,-3ex);
\coordinate (C) at (-9ex,4ex);
\draw (C) circle (0.01ex) node[anchor=east]{\footnotesize{$\tau$}};
\draw[thick,dashed] (-9ex,4ex) -- (9ex,4ex);
\draw[thick] (0,-3ex) -- (-8ex,4ex);
\draw[thick] (0,-3ex) -- (-4ex,4ex);
\draw[thick] (0,-3ex) -- (4ex,4ex);
\draw[thick] (0,-3ex) -- (8ex,4ex);
\filldraw[color=black, fill=black, thick] (-1.2ex,2ex) circle (0.1ex);
\filldraw[color=black, fill=black, thick] (0ex,2ex) circle (0.1ex);
\filldraw[color=black, fill=black, thick] (1.2ex,2ex) circle (0.1ex);
\node at ($(P) + (-0.8ex,0)$) [anchor=east]{\footnotesize{$\lambda^{\text{b}}_{n+2L},\:\bar\tau$}};
\filldraw[color=black, fill=white, thick] (-8.5ex,3.5ex) rectangle (-7.5ex,4.5ex) node[anchor=south]{\footnotesize{$k_1$}};
\filldraw[color=black, fill=white, thick] (-4.5ex,3.5ex) rectangle (-3.5ex,4.5ex) node[anchor=south]{\footnotesize{$k_2$}};
\filldraw[color=black, fill=white, thick] (3.5ex,3.5ex) rectangle (4.5ex,4.5ex) node[anchor=south]{\footnotesize{$k_{n-1}$}};
\filldraw[color=black, fill=white, thick] (7.5ex,3.5ex) rectangle (8.5ex,4.5ex) node[anchor=south]{\footnotesize{$k_n$}};
\draw[thick,scale=3] (0,-1ex)  to[in=-100,out=-140,loop] (0,-1ex);
\draw[thick,scale=3] (0,-1ex)  to[in=-50,out=-90,loop] (0,-1ex);
\draw[thick,scale=3] (0,-1ex)  to[in=30,out=-10,loop] (0,-1ex);
\pgfmathsetmacro{\distanceone}{0.8}
\pgfmathsetmacro{\angleone}{315}
\pgfmathsetmacro{\distancetwo}{0.8}
\pgfmathsetmacro{\angletwo}{330}
\pgfmathsetmacro{\distancethree}{0.8}
\pgfmathsetmacro{\anglethree}{345}
\coordinate (Qone) at ($(P) + (\angleone:\distanceone)$);
\filldraw[color=black, fill=black, thick] (Qone) circle (0.1ex);
\coordinate (Qtwo) at ($(P) + (\angletwo:\distancetwo)$);
\filldraw[color=black, fill=black, thick] (Qtwo) circle (0.1ex) node[anchor=west]{\scriptsize{$L$ loops}};
\coordinate (Qthree) at ($(P) + (\anglethree:\distancethree)$);
\filldraw[color=black, fill=black, thick] (Qthree) circle (0.1ex);
\filldraw[color=white, fill=white] (P) circle (0.8ex);
\filldraw[pattern=north east lines, thick] (P) circle (0.8ex);
}
}
\begin{widetext}
\begin{equation} \label{eq:arbitrary-diagram}
    \Oconnectedloopstwo \!\!\!\!\!\!\!\!\!\!\!= (2\pi)^3 \delta^{(3)} (\boldsymbol{K}) \, 2\,\text{Im}\Bigg\{\frac{\lambda^{\rm b}_{n+2L}}{2^L L! H^4} 
     \int_{-\infty}^{\tau} \frac{\dd{\bar\tau}}{{\bar\tau}^{4}}G_+(\bar\tau,k_1)\cdots G_+(\bar\tau,k_n)\left[\int\frac{\dd[3]{k}}{(2\pi)^3}G_{++}(\bar\tau,\bar\tau,k)\right]^L\!\Bigg\} ,
\end{equation}
\end{widetext}
where $\boldsymbol{K}\equiv {\bk}_1 + \cdots + {\bk}_n$.  Here $G_+(\bar\tau, k)$ represents a bulk-to-boundary propagator connecting the vertex at a time $\bar\tau$ with the boundary at a time $\tau$. On the other hand, $G_{++}(\bar\tau, \bar\tau, k)$ represents a bulk-to-bulk propagator, in this case connecting the vertex at time $\bar\tau$ with itself. Concretely, it is given by
\be
G_{++}(\bar\tau, \bar\tau, k) = f_k(\bar\tau) f^*_k(\bar\tau) .
\ee
In Appendix~\ref{app:feynman}, we summarize the Feynman rules leading to the expressions we are dealing with. The diagram of Eq.~\eqref{eq:arbitrary-diagram} has a symmetry factor $2^L L!$ due to both the symmetry of a single loop ($2^L$) and the interchange of loops ($L!$).

Adding up all the diagrams, we then obtain the following full expression for the $n$-point correlation function:
\begin{flalign} \label{full-correlator-1}
& \!\!\! \langle \varphi_{{\bk}_1} \cdots\, \varphi_{{\bk}_n} \rangle (\tau) = (2 \pi)^3 \delta(\boldsymbol{K}) \sum_{L} \frac{\lambda_{n+2L}^{b}}{H^4 L!}  \notag \\ &  \times  2\,\text{Im}\Bigg\{ \! \int_{-\infty}^{\tau} \!\! \frac{\dd{\bar\tau}}{{\bar\tau}^{4}}G_+(\bar\tau,k_1)\dotso G_+(\bar\tau,k_n) \left( \frac{\sigma_{\rm tot}^2}{2} \right)^L \! \Bigg\} ,
\end{flalign}
where we have defined $\sigma_{\rm tot}^2\equiv  \int\frac{\dd[3]{k}}{(2\pi)^3}G_{++}(\bar\tau,\bar\tau,k)$, which, after integrating the solid angle, becomes 
\begin{equation} \label{sigma-tot-2-def}
\sigma_{\rm tot}^2  = \frac{1}{2 \pi^2} \int_0^{\infty} \frac{\dd{k}}{k} k^3 G_{++}(\bar\tau,\bar\tau,k) .
\end{equation}
This quantity measures the integral contribution of a single loop inside the $\bar\tau$-integral. At first sight this quantity appears to be time-dependent, implying that the overall time-dependence of the integrand in Eq.~\eqref{eq:arbitrary-diagram} gets more complicated with more and more loops. However, $k^3 G_{++} (\bar \tau, \bar \tau , k)$ is a function of the combination $k \tau$. Thus, after performing the change of variables $k \to p = k / a(\tau)=-Hk\tau $ from comoving momentum $k$ to physical momentum $p$, one trivially finds that $\sigma_{\rm tot}^2$ is a constant. For instance, in the particular case of a massless field in  de Sitter spacetime, one finds
\begin{equation} \label{sigma-tot}
\sigma_{\rm tot}^2  =\frac{H^2}{4 \pi^2} \int_0^{\infty}  \frac{\dd p}{p}  \left( 1 + \frac{p^2}{H^2} \right) .
\end{equation}
Alternatively, for a massive field of mass $m$, one finds
\begin{equation}
\sigma_{\rm tot}^2  = \frac{1}{16\pi H} \int_0^{\infty} \frac{\dd p}{p} p^{3} \left|H_\nu^{(1)}\left(\frac{p}{H}\right)\right|^2 .
\end{equation}
Consequently, the quantity $\sigma_{\rm tot}^{2 L}$ can be pulled out of the $\bar\tau$-integral of Eq.~\eqref{full-correlator-1} to become an overall constant factor. In this way, Eq.~\eqref{full-correlator-1} becomes
\begin{flalign} \label{full-correlator-2}
& \langle \varphi_{{\bk}_1} \cdots \,\varphi_{{\bk}_n} \rangle (\tau) = (2 \pi)^3 \delta(\boldsymbol{K}) \frac{\lambda_n^{\rm obs}}{H^4}  \notag \\ & \;\;\;\;\;\;  \times  2\,\text{Im}\Bigg\{  \int_{-\infty}^{\tau} \frac{\dd{\bar\tau}}{{\bar\tau}^{4}}G_+(\bar\tau,k_1)\dotso G_+(\bar\tau,k_n) \Bigg\} ,
\end{flalign}
where $\lambda_n^{\rm obs}$ is the observable coupling defined as
\be \label{lambda_obs-def}
\lambda_n^{\rm obs} \equiv \sum_{L=0}^\infty \frac{\lambda_{n+2L}^{\rm b}}{L!} \left( \frac{\sigma_{\rm tot}^2}{2} \right)^L .
\ee
Notice that Eq.~\eqref{full-correlator-2} is equivalent to a tree-level diagram with a single $n$-legged vertex of strength $\lambda_n^{\rm obs}$. Of course, despite being a constant, $\sigma_{\rm tot}^{2}$ is clearly divergent due to both the upper and the lower bound of the integral (e.g., in Eq.~\eqref{sigma-tot}). On the other hand, Eq.~\eqref{lambda_obs-def} must be finite. In the following two sections, we do not dwell on the specific form of $\sigma_{\rm tot}^2$, and instead examine how to deal with the divergence it implies.

At this point, some readers may object that the time independence of $\sigma_{\rm tot}^{2}$ has been enforced by the change of variables $k \to p = k / a(\tau)$. In Sec.~\ref{sec:dim-reg}, we will compute this quantity using dimensional regularization. There, we shall confirm that $\sigma_{\rm tot}^2$ is indeed constant, with time-independent divergent pieces that can be absorbed by counterterms in the way we proceed to explain in the following section.

Before moving on to the next section, recall that the couplings $\lambda_n^{\rm b}$ are the Taylor coefficients determining the bare potential $\V_{\rm b}(\varphi)$ through the expansion in Eq.~\eqref{eqn:Vb}. Eq.~\eqref{lambda_obs-def} may be seen as defining new Taylor coefficients $\lambda_n^{\rm obs}$ for an observable potential $\V_{\rm obs}(\varphi)$ by writing
\begin{equation}
    \V_{\rm obs}(\varphi)=\sum_{n=0}^\infty\frac{\lambda_n^{\rm obs}}{n!}\varphi^n .
    \label{eqn:Vobs}
\end{equation}
One may then ask how the potentials $\V_{\rm obs}$ and $\V_{\rm b}$ are related. It turns out that they are related via a Weierstrass transformation given by
\begin{equation} \label{weierstrass-obs-b}
    \V_{\rm obs} (\varphi) =  e^{ \frac{1}{2} \sigma_{\rm tot}^2 \frac{\partial^2}{\partial \varphi^2}}  \V_{\rm b} (\varphi) ,
\end{equation}
which is equivalent to Eq.~\eqref{lambda_obs-def}. This result already offers hints on how to deal with the divergence implied by $\sigma_{\rm tot}^2$. We will re-obtain this result perturbatively in the next section.

\section{Basic strategy} \label{sec:basic}

Let us split the bare coupling $\lambda_n^{\rm b}$ into a renormalized coupling $\lambda_n$ and a counterterm $\lambda_n^{\rm ct}$ as 
\begin{equation}
    \lambda_n^{\rm b} = \lambda_n + \lambda_n^{\rm ct}.
    \label{eqn:lambdab}
\end{equation}
More generally, we should also consider normalizing the field $\varphi$ and splitting the field normalization analogously. However, given that the diagrams we are examining have loops that carry no external momenta, this is not necessary (i.e., we do not need to worry about renormalization of the kinetic term of the bare theory). Next, given that we are dealing with diagrams with an arbitrary number of loops, we may further split $ \lambda_n^{\rm ct}$ into terms of order $L$, needed to cancel divergences coming from $L$-loop diagrams. Thus we write:
\begin{equation}
  \lambda_n^{\rm ct} = \lambda_n^{(1)} + \lambda_n^{(2)} + \lambda_n^{(3)} + \cdots .
\end{equation}
Notationally, we may thus simply write 
\begin{equation} \label{lambda_b-lambda_L}
    \lambda_n^{\rm b} = \sum_{L=0}^{\infty} \lambda_n^{(L)} ,
\end{equation}
where $\lambda_n^{(0)} =  \lambda_n$ is just the renormalized coupling already introduced in Eq.~\eqref{eqn:lambdab}.

Having defined this splitting, we must re-work the computation of an $n$-point correlation function by reorganizing diagrams in terms of the couplings $ \lambda_n^{(L)}$. To be precise, a diagram with $L_1$ loops proportional to a counterterm with coupling $\lambda_{n+2 L_1}^{(L_2)}$ must be regarded as of order $L= L_1 + L_2$. The full correlation function can then be written as
\be \label{full-sum-L}
\langle \varphi_{{\bk}_1} \cdots \,\varphi_{{\bk}_n} \rangle (\tau) = \sum_{L=0}^\infty \langle \varphi_{{\bk}_1} \cdots \,\varphi_{{\bk}_n} \rangle^{(L)} (\tau) ,
\ee
where $\langle \varphi_{{\bk}_1} \cdots \,\varphi_{{\bk}_n} \rangle^{(L)} (\tau) $ contains every contribution, including loops and counterterms, of order $L$. Diagramatically, this is given by:
\begin{widetext}
\def\Oconnectedloopsren{\tikz[baseline=-1.4ex]{
\coordinate (P) at (0,-3ex);
\coordinate (C) at (-9ex,4ex);
\draw (C) circle (0.01ex) node[anchor=east]{\footnotesize{$\tau$}};
\draw[thick,dashed] (-9ex,4ex) -- (9ex,4ex);
\draw[thick] (0,-3ex) -- (-8ex,4ex);
\draw[thick] (0,-3ex) -- (-4ex,4ex);
\draw[thick] (0,-3ex) -- (4ex,4ex);
\draw[thick] (0,-3ex) -- (8ex,4ex);
\filldraw[color=black, fill=black, thick] (-1.2ex,2ex) circle (0.1ex);
\filldraw[color=black, fill=black, thick] (0ex,2ex) circle (0.1ex);
\filldraw[color=black, fill=black, thick] (1.2ex,2ex) circle (0.1ex);
\node at ($(P) + (-0.8ex,0)$) [anchor=east]{\footnotesize{$\lambda^{(0)}_{n+2L},\:\bar\tau$}};
\filldraw[color=black, fill=white, thick] (-8.5ex,3.5ex) rectangle (-7.5ex,4.5ex) node[anchor=south]{\footnotesize{$k_1$}};
\filldraw[color=black, fill=white, thick] (-4.5ex,3.5ex) rectangle (-3.5ex,4.5ex) node[anchor=south]{\footnotesize{$k_2$}};
\filldraw[color=black, fill=white, thick] (3.5ex,3.5ex) rectangle (4.5ex,4.5ex) node[anchor=south]{\footnotesize{$k_{n-1}$}};
\filldraw[color=black, fill=white, thick] (7.5ex,3.5ex) rectangle (8.5ex,4.5ex) node[anchor=south]{\footnotesize{$k_n$}};
\draw[thick,scale=3] (0,-1ex)  to[in=-100,out=-140,loop] (0,-1ex);
\draw[thick,scale=3] (0,-1ex)  to[in=-50,out=-90,loop] (0,-1ex);
\draw[thick,scale=3] (0,-1ex)  to[in=30,out=-10,loop] (0,-1ex);
\pgfmathsetmacro{\distanceone}{0.8}
\pgfmathsetmacro{\angleone}{315}
\pgfmathsetmacro{\distancetwo}{0.8}
\pgfmathsetmacro{\angletwo}{330}
\pgfmathsetmacro{\distancethree}{0.8}
\pgfmathsetmacro{\anglethree}{345}
\coordinate (Qone) at ($(P) + (\angleone:\distanceone)$);
\filldraw[color=black, fill=black, thick] (Qone) circle (0.1ex);
\coordinate (Qtwo) at ($(P) + (\angletwo:\distancetwo)$);
\filldraw[color=black, fill=black, thick] (Qtwo) circle (0.1ex) node[anchor=west]{\scriptsize{$L$ loops}};
\coordinate (Qthree) at ($(P) + (\anglethree:\distancethree)$);
\filldraw[color=black, fill=black, thick] (Qthree) circle (0.1ex);
\filldraw[color=white, fill=white] (P) circle (0.8ex);
\filldraw[pattern=north east lines, thick] (P) circle (0.8ex);
}
}
\def\OconnectedloopsLminuoneCountertermVtwo{\tikz[baseline=-1.4ex]{
\coordinate (P) at (0,-3ex);
\coordinate (C) at (-9ex,4ex);
\draw (C) circle (0.01ex) node[anchor=east]{\footnotesize{$\tau$}};
\draw[thick,dashed] (-9ex,4ex) -- (9ex,4ex);
\draw[thick] (0,-3ex) -- (-8ex,4ex);
\draw[thick] (0,-3ex) -- (-4ex,4ex);
\draw[thick] (0,-3ex) -- (4ex,4ex);
\draw[thick] (0,-3ex) -- (8ex,4ex);
\filldraw[color=black, fill=black, thick] (-1.2ex,2ex) circle (0.1ex);
\filldraw[color=black, fill=black, thick] (0ex,2ex) circle (0.1ex);
\filldraw[color=black, fill=black, thick] (1.2ex,2ex) circle (0.1ex);
\node at ($(P) + (-0.5ex,0)$) [anchor=east]{\footnotesize{$\lambda^{(1)}_{n+2(L-1)},\:\bar\tau$}};
\filldraw[color=black, fill=white, thick] (-8.5ex,3.5ex) rectangle (-7.5ex,4.5ex) node[anchor=south]{\footnotesize{$k_1$}};
\filldraw[color=black, fill=white, thick] (-4.5ex,3.5ex) rectangle (-3.5ex,4.5ex) node[anchor=south]{\footnotesize{$k_2$}};
\filldraw[color=black, fill=white, thick] (3.5ex,3.5ex) rectangle (4.5ex,4.5ex) node[anchor=south]{\footnotesize{$k_{n-1}$}};
\filldraw[color=black, fill=white, thick] (7.5ex,3.5ex) rectangle (8.5ex,4.5ex) node[anchor=south]{\footnotesize{$k_n$}};
\draw[thick,scale=3] (0,-1ex)  to[in=-50,out=-90,loop] (0,-1ex);
\draw[thick,scale=3] (0,-1ex)  to[in=30,out=-10,loop] (0,-1ex);
\pgfmathsetmacro{\distanceone}{0.8}
\pgfmathsetmacro{\angleone}{315}
\pgfmathsetmacro{\distancetwo}{0.8}
\pgfmathsetmacro{\angletwo}{330}
\pgfmathsetmacro{\distancethree}{0.8}
\pgfmathsetmacro{\anglethree}{345}
\coordinate (Qone) at ($(P) + (\angleone:\distanceone)$);
\filldraw[color=black, fill=black, thick] (Qone) circle (0.1ex);
\coordinate (Qtwo) at ($(P) + (\angletwo:\distancetwo)$);
\filldraw[color=black, fill=black, thick] (Qtwo) circle (0.1ex) node[anchor=west]{\scriptsize{$(L-1)$ loops}};
\coordinate (Qthree) at ($(P) + (\anglethree:\distancethree)$);
\filldraw[color=black, fill=black, thick] (Qthree) circle (0.1ex);
\filldraw[color=white, fill=white] (P) circle (0.8ex);
\filldraw[pattern=north east lines, thick] (P) circle (0.8ex);
}
}
\def\NlegsCountertermOrderLVtwo{\tikz[baseline=-1.4ex]{
\coordinate (P) at (0,-3ex);
\coordinate (C) at (-9ex,4ex);
\draw (C) circle (0.01ex) node[anchor=east]{\footnotesize{$\tau$}};
\draw[thick,dashed] (-9ex,4ex) -- (9ex,4ex);
\draw[thick] (0,-3ex) -- (-8ex,4ex);
\draw[thick] (0,-3ex) -- (-4ex,4ex);
\draw[thick] (0,-3ex) -- (4ex,4ex);
\draw[thick] (0,-3ex) -- (8ex,4ex);
\filldraw[color=black, fill=black, thick] (-1.2ex,2ex) circle (0.1ex);
\filldraw[color=black, fill=black, thick] (0ex,2ex) circle (0.1ex);
\filldraw[color=black, fill=black, thick] (1.2ex,2ex) circle (0.1ex);
\node at ($(P) + (-0.5ex,0)$) [anchor=east]{\footnotesize{$\lambda_{n}^{(L)}\!,\:\bar\tau$}};
\filldraw[color=black, fill=white, thick] (-8.5ex,3.5ex) rectangle (-7.5ex,4.5ex) node[anchor=south]{\footnotesize{$k_1$}};
\filldraw[color=black, fill=white, thick] (-4.5ex,3.5ex) rectangle (-3.5ex,4.5ex) node[anchor=south]{\footnotesize{$k_2$}};
\filldraw[color=black, fill=white, thick] (3.5ex,3.5ex) rectangle (4.5ex,4.5ex) node[anchor=south]{\footnotesize{$k_{n-1}$}};
\filldraw[color=black, fill=white, thick] (7.5ex,3.5ex) rectangle (8.5ex,4.5ex) node[anchor=south]{\footnotesize{$k_n$}};
\filldraw[color=white, fill=white] (P) circle (0.8ex);
\filldraw[pattern=north east lines, thick] (P) circle (0.8ex);
}
}
\be
\langle \varphi_{{\bk}_1} \cdots \,\varphi_{{\bk}_n} \rangle^{(L)} (\tau) = \Oconnectedloopsren + \OconnectedloopsLminuoneCountertermVtwo \!\!\!\!\!\!\!\!\!\!+ \, \cdots \, + \NlegsCountertermOrderLVtwo .
\ee
\end{widetext}
Working out these diagrams explicitly, one finds that $\langle \varphi_{{\bk}_1} \cdots\, \varphi_{{\bk}_n} \rangle^{(L)} (\tau)$ is given by
\begin{flalign} 
& \!\!\! \langle \varphi_{{\bk}_1} \cdots\, \varphi_{{\bk}_n} \rangle^{(L)} (\tau) =  \frac{(2 \pi)^3}{H^4}  \delta(\boldsymbol{K})  \sum_{s=0}^{L} \frac{1}{s!} \left( \frac{\sigma_{\rm tot}^2}{2} \right)^{s}  \lambda^{(L-s)}_{n+2s} \notag \\ & \;\;\;\;\;\;\;\;\;\;\;\;\;\;\;\;\; \times  2\,\text{Im}\Bigg\{ \! \int_{-\infty}^{\tau} \!\! \frac{\dd{\bar\tau}}{{\bar\tau}^{4}}G_+(\bar\tau,k_1)\dotso G_+(\bar\tau,k_n)  \Bigg\} .
\end{flalign}
Substituting this expression back into Eq.~\eqref{full-sum-L} and comparing the result to Eq.~\eqref{full-correlator-2}, one sees that the observable coupling $\lambda_n^{\rm obs}$ is simply given by
\begin{equation} \label{lambda-obs-bare}
\lambda_n^{\rm obs} \equiv \sum_{L=0}^{\infty}  \sum_{s=0}^{L} \frac{1}{s!} \left( \frac{\sigma_{\rm tot}^2}{2} \right)^{s}  \lambda^{(L-s)}_{n+2s} \; . 
\end{equation}
This is, of course, nothing but Eq.~\eqref{lambda_obs-def} after substituting in the expansion of Eq.~\eqref{lambda_b-lambda_L}.

The full expression must be finite, and so now the challenge is to determine the appropriate values for the counterterm couplings $\lambda_n^{(L)}$ (with $L>0$) in terms of the renormalized couplings $\lambda_n^{(0)} = \lambda_n$. As a first approach, let us require that the counterterms $\lambda_n^{(L)}$ exactly cancel the divergent constant $\sigma_{\rm tot}^2$ order by order in the loop expansion. This approach corresponds to the on-shell scheme usually encountered in quantum field theory, and it requires the condition $\langle \varphi_{{\bk}_1} \cdots \,\varphi_{{\bk}_n} \rangle^{(L)} (\tau) = 0$  for all $L > 0$. In turn, this condition implies the following infinite set of algebraic equations:
\begin{equation} \label{algebraic-system}
\sum_{s=0}^{L} \frac{1}{s!} \left( \frac{\sigma_{\rm tot}^2}{2} \right)^{s}  \lambda^{(L-s)}_{n+2s}  = 0 , \qquad {\rm for \,\, all \,\,} L > 0 .
\end{equation}
Recall that we are working with units where $\hbar = 1$. If we had chosen to work with $\hbar \neq 1$, then this condition would ensure that each contribution to $\lambda_n^{\rm obs}$ of order $\hbar^L$ cancels out. The solution to this set of equations turns out to be (see Appendix~\ref{app:proof}):
\begin{equation} \label{solution-div-1}
    \lambda_n^{(L)} = (-1)^L \frac{1}{L!} \left( \frac{\sigma_{\rm tot}^2}{2} \right)^L \lambda_{n+2L}^{(0)} ,
\end{equation}
which may be proved by induction. With this choice of counterterms, the full expression for the $n$-point correlation function in Eq.~\eqref{full-correlator-2} becomes finite, with $\lambda_n^{\rm obs} = \lambda_n^{(0)} = \lambda_n$. This result, together with the solution in Eq.~\eqref{solution-div-1}, allows us to express the original divergent bare couplings $\lambda_n^{\rm b}$ in terms of the observable couplings $\lambda_n^{\rm obs}$ as:
\begin{equation} \label{bare-lambda-ren-lambda}
\lambda_n^{\rm b} = \sum_{L=0}^{\infty} (-1)^L \frac{1}{L!} \left( \frac{\sigma_{\rm tot}^2}{2} \right)^L  \lambda_{n+2L}^{\rm obs} .
\end{equation}

The previous expression is similar in form to Eq.~\eqref{lambda_obs-def}. In fact, recall that we used Eq.~\eqref{lambda_obs-def} to derive Eq.~\eqref{weierstrass-obs-b}, which states that $\mathcal V_{\rm obs} (\varphi)$ is a Weierstrass transformation of $\mathcal V_{\rm b} (\varphi)$. Following exactly the same reasoning, we can now see that Eq.~\eqref{bare-lambda-ren-lambda} leads to the inverse Weierstrass transformation 
\begin{equation} \label{weierstrass-b-obs}
    \V_{\rm b} (\varphi) =  e^{- \frac{1}{2} \sigma_{\rm tot}^2 \frac{\partial^2}{\partial \varphi^2}}  \V_{\rm obs} (\varphi) .
\end{equation}
In this way, observables, up to order $\lambda$, will depend on the observable potential $\V_{\rm obs} (\varphi)$, and not on the formally infinite potential $\V_{\rm b} (\varphi) $. This agrees with the results of Refs.~\cite{Chen:2018brw,Palma:2023idj}.
 
\section{Renormalization scheme} \label{sec:ren-scheem} 
 
There must exist some arbitrariness in the choice of the counterterm couplings $\lambda_n^{(L)}$, for $L>0$, as usual in standard perturbation theory. For instance, in dimensional regularization, $\sigma_{\rm tot}^2$ might have a divergent piece proportional to $\delta^{-1}$ (where $\delta$ parametrizes the departure from $3$ spatial dimensions), together with a finite part. Thus, if we aim for a minimal subtraction scheme, we should tune the counterterms $\lambda_n^{(L)}$ to be proportional to $\delta^{-L}$ in order to capture only the divergent contributions. 

To proceed, let us split $\sigma_{\rm tot}^2$ into two terms as $\sigma_{\rm tot}^2 = \sigma_{0}^2 + \sigma_{\infty}^2$, and choose  the counterterms $\lambda_n^{(L)}$ in such a way that they cancel out the divergent piece, say, $\sigma_{\infty}^2$. In this case, the effective coupling can be written as
 \begin{equation} \label{lambda-eff-a-b}
\lambda_n^{\rm obs} = \sum_{L=0}^{\infty}  \sum_{s=0}^{L} \frac{1}{s!} \left(\frac{1}{2}\sigma_{0}^2 + \frac{1}{2}\sigma_{\infty}^2 \right)^{s}  \lambda^{(L-s)}_{n+2s} .
\end{equation}
Now, motivated by the solution in Eq.~\eqref{solution-div-1}, let us guess that the correct choice for the counterterm couplings $\lambda_n^{(L)}$ (for $L >0$) is given by
\begin{equation} \label{solution-div-2}
    \lambda_n^{(L)} = (-1)^L \frac{1}{L!} \left( \frac{1}{2}\sigma_{\infty}^{2}  \right)^L \lambda_{n+2L}^{(0)} .
\end{equation}

By plugging this ansatz back into Eq.~\eqref{lambda-eff-a-b}, we are led to
\bea
\lambda_n^{\rm obs} &=& \sum_{L=0}^{\infty}  \frac{\lambda^{(0)}_{n+2L}}{L!} \sum_{s=0}^{L}   \frac{L!}{s! (L-s)!} \nn \\ &&  \times \left(- \frac{1}{2}\sigma_{\infty}^2 \right)^{L-s}\left( \frac{1}{2}\sigma_{0}^2 + \frac{1}{2}\sigma_{\infty}^2 \right)^{s} .
\eea
Then, by recognizing the general formula $(x + y)^L = \sum_{s=0}^{L} \frac{L!}{s! (L-s)!} x^{L-s} y^s$, we see that:
\begin{equation} \label{lambda-obs-a}
\lambda_n^{\rm obs} = \sum_{L=0}^{\infty}  \frac{\lambda^{(0)}_{n+2L}}{L!} \left( \frac{1}{2} \sigma_{0}^{2} \right)^L .
\end{equation}
Thus, with the choice in Eq.~\eqref{solution-div-2}, one can remove the specific divergent part of $\sigma_{\rm tot}^2$ and retain the finite contribution. 

Let us now define the renormalized potential $\V_{\rm ren}(\varphi)$ as the resulting resummation of the Taylor coefficients $\lambda_{n}^{(0)}$: 
\begin{equation}
    \V_{\rm ren} (\varphi) \equiv \sum_{n=0}^\infty\frac{\lambda^{(0)}_n}{n!}\varphi^n .
    \label{eqn:Veff}
\end{equation}
Notice that in Sec.~\ref{sec:basic}, $\V_{\rm ren}(\varphi)$ and $\V_{\rm obs}(\varphi)$ coincided because $\lambda_n^{\rm obs} = \lambda_n^{(0)}$. Here, given that $\sigma_{0}^2  \neq 0$, Eq.~\eqref{lambda-obs-a} implies that  
\begin{equation} \label{V_obs-V_ren}
    \V_{\rm obs} (\varphi) =  e^{\frac{\sigma_{0}^2}{2} \frac{\partial^2}{\partial \varphi^2}}  \V_{\rm ren} (\varphi) .
\end{equation}
On the other hand, from Eq.~\eqref{solution-div-2}, we see that
\begin{equation} \label{V_b-V_ren}
    \V_{\rm b} (\varphi) =  e^{-\frac{\sigma_{\infty}^2}{2} \frac{\partial^2}{\partial \varphi^2}}  \V_{\rm ren} (\varphi) .
\end{equation}
Both Eqs.~\eqref{V_obs-V_ren} and \eqref{V_b-V_ren} directly imply $\V_{\rm obs} (\varphi) =  e^{\frac{\sigma_{\rm tot}^2}{2} \frac{\partial^2}{\partial \varphi^2}}  \V_{\rm b} (\varphi)$, which is the result previously derived at the end of Sec.~\ref{sec:correlators}. Thus, the arbitrariness in the choice of the counterterms $\lambda_n^{(L)}$ is reflected in the arbitrary parameter $\sigma_{0}^2$ connecting the observable potential $\V_{\rm obs} (\varphi)$ and the renormalized potential $\V_{\rm ren} (\varphi)$. As usual, the observable potential $\V_{\rm obs} (\varphi)$ cannot depend on this arbitrary parameter, and, as such, the renormalized potential $\V_{\rm ren} (\varphi)$ must depend on $\sigma_{0}^2$ in such a way that it cancels out after the Weierstrass operator $\exp \left\{ \frac{1}{2}  \sigma_{0}^2 \partial_\varphi^2 \right\} $ acts on it. In the following section, we discuss the renormalization flow that the parameter $\sigma_0^2$ implies.

\vspace{0.5cm}
\section{Dimensional regularization} \label{sec:dim-reg}

To put our approach in more concrete terms, let us analyze the divergent integrals within dimensional regularization. To proceed, we must first extend some relevant quantities from $d=3$ to an arbitrary spatial dimension $d=3+\delta$. To start with, notice that in $3$ dimensions the couplings $\lambda_n$ have mass-dimension $[\lambda_n]=4 - n$. We may conveniently keep $[\lambda_n]$ fixed in arbitrary spatial dimension $d = 3 + \delta$ by performing the following replacement:
\begin{equation}
    \lambda_n\to \lambda_n\tilde{\mu}^{\delta(1-n/2)}\,,
\end{equation}
where $\tilde{\mu}$ is the usual renormalization mass scale. The field has mass-dimension $[\varphi]=1 + \delta/2$, which in turn implies that $f_k(\tau)$ has a fixed mass-dimension of $-1/2$. For this reason, it is convenient to redefine $\sigma_{\rm tot}^2$ in a way that its mass-dimension remains fixed to $2$. The appropriate redefinition anticipates the relation between the couplings $\lambda_n$ and $\sigma_{\rm tot}^2$ presented in Eq.~\eqref{lambda-obs-bare}:
\begin{equation} \label{sigma-delta}
\sigma_{\rm tot}^2  \equiv  \frac{1}{\tilde \mu^{\delta}} \int\frac{\dd[3 + \delta]{k}}{(2\pi)^{3+\delta}} f_k(\tau) f^*_{k} (\tau) .
\end{equation}

The equation of motion respected by $f_k(\tau)$ in $d = 3 + \delta$ dimensions is given by:
\be \label{f-eq-delta}
f_k'' - \frac{2+\delta}{\tau} f_k' + \left( k^2 + \frac{m^2}{H^2 \tau^2} \right) f_k = 0 .
\ee
The solution to this equation with appropriate Bunch-Davies initial conditions leads to~\cite{Senatore:2009cf}
\begin{equation} \label{fk-delta}
    f_k(\tau) = \frac{i\sqrt{\pi}H^{1+\delta/2}}{2} (-\tau)^{(3+\delta)/2} H_\nu^{(1)}(-k\tau),
\end{equation}
where $H_{\nu}^{(1)}(x)$ is a Hankel function of the first kind with parameter $\nu = \sqrt{(3 + \delta)^2/4 - m^2/H^2}$. We may now compute $\sigma_{\rm tot}^2$ by substituting Eq.~\eqref{fk-delta} back into Eq.~\eqref{sigma-delta}; one finds:
\begin{equation}  \label{sigma-tot-x}
\sigma_{\rm tot}^2 =  \frac{\pi^{-(1+\delta)/2} H^{2+\delta}}{16 \,   \Gamma \! \left(\frac{3+\delta}{2}\right) ( 2 \tilde \mu)^{\delta}}  \int_0^{\infty} \frac{\dd x}{x} x^{3+\delta}  \left| H_\nu^{(1)}(x) \right|^2 ,
\end{equation}
where we made the change of variables $x = - k \tau$. Notice that, just as discussed in Sec.~\ref{sec:correlators}, the loop factor $\sigma_{\rm tot}^2$ is now explicitly time independent. What remains now is to integrate this expression~\cite{wolfram}, which is possible as long as $ \delta < -2$ and $\delta - 2\nu > - 3$. Where these conditions are satisfied, the integral results in:
\begin{widetext}
\begin{equation} \label{sigma-delta-mu}
\sigma_{\rm tot}^2  =  \frac{\pi^{- \delta/2}}{16 \, \Gamma \! \left[\frac{3 + \delta}{2}\right]} \frac{ H^{2+\delta}}{ (2 \tilde \mu)^{\delta}}   \frac{  \cos[ \pi \nu ] \csc[ \pi (3+\delta) /2 ] \csc[ \pi ((3+\delta)/2 - \nu )  ] \csc [ \pi ((3+\delta)/2 + \nu)  ] \Gamma[ -( 2 + \delta)/2 ]}{   \Gamma[  - (1 + \delta) /2 ] \Gamma[ -  (1 + \delta)/2 - \nu ] \Gamma[  - (1 + \delta)/2 + \nu ]} .
\end{equation}
\end{widetext}

We may now analytically continue this result for all values of $\delta$. The quantity $\sigma_{\rm tot}^2$ is a complicated function of $\delta$ and $\nu$ in the vicinity of $(\delta , \nu) =(0, 3/2)$, and the divergent nature of $\sigma_{\rm tot}^2$ depends on how we choose to approach this point. One way to deal with this complication, as explored in Refs.~\cite{Melville:2021lst,Lee:2023jby}, consists in fixing the mass parameter $m^2$ appearing in Eq.~\eqref{f-eq-delta} as $m^2  = m_{\rm ph}^2 + H^2(d^2- 9)/4$, with $m_{\rm ph}$ identified as the physical mass of $\varphi$ in the free-theory case, where $\mathcal V (\varphi) = 0$. This choice ensures that $\nu = \sqrt{9/4 - m_{\rm ph}^2 / H^2}$ for all values of $d$, and the theory remains scale invariant in the vicinity of $d=3$ for the particular case of a massless field ($m_{\rm ph} = 0$). With this choice, it is then possible to verify that 
\be
\sigma_{\rm tot}^2 \to 0  \qquad {\rm for} \qquad \nu \to 3/2 ,
\ee
which corresponds to the limit $m_{\rm ph} \to 0$, independent of the value of $\delta$~\cite{Lee:2023jby, Creminelli2024}. This is simply because the $x$-integral of Eq.~\eqref{sigma-tot-x} becomes $\int_0^{\infty} \dd x x^{\delta} (1 + x^2)$, which, being the integral of a polynomial, vanishes in dimensional regularization~\cite{Collins_1984}.\footnote{Note that dimensional regularization only captures logarithmic divergencies, while it is blind to power-law ones~\cite{Collins_1984,Bonanno:2020bil}.}

Instead of fixing the mass parameter $m^2$ in the way just described, we could simply expand $\sigma_{\rm tot}^2$ first in terms of $m^2/H^2$ and then in terms of $\delta$. One then finds:
\be
\!\sigma_{\rm tot}^2\! = - \frac{H^2}{4 \pi^{2} \delta} + \frac{3 H^4}{8  \pi^{2} m^2} + \frac{H^2}{4 \pi^{2}}    \ln \left( \frac{\mu}{H}  \right)   + \mathcal O \qty(\delta, \frac{m^2}{H^2}),
\ee
where we have redefined $\mu \equiv \sqrt{4 \pi} e^{(\gamma_{\rm E}/2  - 7/4 )} \tilde \mu$, with $\gamma_{\rm E}$ the Euler-Mascheroni constant. The previous result explicitly displays two divergent contributions, one from $\delta \to 0$, which may be identified with an ultraviolet divergence, and a second one, from $m^2 / H^2 \to 0$, which embodies the infrared divergent part. Of course, these divergences are not a problem as we can now proceed to minimally subtract them in the way explained in the previous section. We end up finding a finite contribution $\sigma_0^2$ for loop integrals, given by:
\be \label{sigma-0-mu-v1}
\sigma_0^2 = \frac{H^2}{4 \pi^{2}}    \ln \left( \frac{\mu}{H}  \right)  .
\ee
The arbitrariness of the subtraction, as usual, is reflected in the dependence of the remaining finite part on the renormalization scale $\mu$. The fact that the loops discussed here have no physical effect aligns with the considerations of Refs.~\cite{Lee:2023jby,Creminelli2024}, which also perform dimensional regularization.

As commented in Sec.~\ref{sec:ren-scheem}, the observable potential (i.e., the potential defining the couplings appearing in observable correlation functions) must be independent of $\mu$. Thus, from Eq.~\eqref{sigma-0-mu-v1}, together with Eq.~\eqref{V_obs-V_ren}, we see that the renormalized potential must satisfy the following renormalization group flow equation:
\be \label{flow}
  \mu \frac{\partial}{\partial \mu}  \V_{\rm ren} (\varphi) = - \frac{H^2}{8 \pi^{2}}    \V_{\rm ren}'' (\varphi) .
\ee
One may put this result in concrete terms by considering an axionic potential $\mathcal V_{\rm ren} (\varphi) =  - A \cos ( \varphi/f )$, with $f$ the axion decay constant. Then, from Eq.~\eqref{V_obs-V_ren}, it follows that the observable potential is read as $\mathcal V_{\rm obs} (\varphi) =  - A_{\rm obs} \cos (\varphi/f )$ with $A_{\rm obs}$ and $A$ related as
\be
A_{\rm obs} = A (\mu) e^{- \frac{1}{2} \frac{\sigma_0^2 (\mu)}{f^2} } .
\ee
Because $A_{\rm obs}$ cannot depend on the renormalization scale, we see that $A (\mu) \propto \exp [  \frac{H^2}{8 \pi^{2} f^2} \ln \qty(\frac{\mu}{H}) ]$. On the other hand, the decay coupling constant $f$ does not receive loop corrections. 

Given that observables to order $\lambda$ do not have external momenta running through loops, we do not see combinations such as $\ln \left(k/\mu \right)$, and so the renormalization scale $\mu$ is not a useful quantity allowing one to assess the flow of momenta through couplings. Nevertheless, in the next section, we will encounter similar expressions involving cutoff momenta that will allow us to assess the validity of perturbation theory of order $\lambda$ at different scales.

\section{Wilsonian approach} \label{sec:Wilsonian}

Let us consider the problem of computing correlation functions within an effective field theory point of view. Our starting point is an effective field theory defined over a range of physical momenta bounded by infrared and ultraviolet cutoff scales $\Lambda_{\rm IR}$ and $\Lambda_{\rm UV}$, respectively
\begin{flalign} \label{action-2}
    & S_{\rm eff} (\Lambda_{\rm IR} , \Lambda_{\rm UV}) \notag \\ &\;\;\;\;\;  = \int  \dd[3]{x}\dd{\tau}a^4\left[\frac{\left(\varphi'\right)^2}{2a^2}-\frac{(\nabla\varphi)^2}{2a^2} -   \V_{\rm eff}(\varphi)    \right].
\end{flalign}
Wilsonian coefficients in $S_{\rm eff}$ incorporate the integration of momenta below $\Lambda_{\rm IR}$ and above $\Lambda_{\rm UV}$, to all order with respect to both vertices and loops. As a consequence, the computation of correlation functions, including loop corrections, must be finite. By writing $\V_{\rm eff} (\varphi) = \sum_{n} \frac{\lambda^{\rm eff}_n}{n!} \varphi^n$, a direct computation to first order with respect to $\lambda^{\rm eff}_n$ leads to\footnote{Strictly speaking, the bulk-to-boundary propagators appearing in~(\ref{obs-n-pt}) should vanish for momenta $k_i |\bar \tau| < \Lambda_{\rm UV}$. However, this effect can be disregarded as the time integral quickly converges for momenta $k_i |\bar \tau| < H$, thanks to the oscillatory nature of the mode functions.}
\begin{flalign} \label{obs-n-pt}
& \!\!\! \langle \varphi_{{\bk}_1} \cdots\, \varphi_{{\bk}_n} \rangle (\tau)  \notag \\ & \,\,= \frac{\lambda_n^{\rm obs}}{H^4} \, 2\,\text{Im}\Bigg\{ \int_{-\infty}^{\tau} \frac{\dd{\bar\tau}}{{\bar\tau}^{4}}G_+(\bar\tau,k_1)\dotso G_+(\bar\tau,k_n) \Bigg\} ,
\end{flalign}
where $\lambda_n^{\rm obs}$ is found to be:
\be \label{obs-eff}
\lambda^{\rm obs}_{n} = \sum_{L=0}^{\infty} \frac{1}{L!} \lambda^{\rm eff}_{n + 2 L}  \left[ \frac{1}{2} \sigma^2 (\Lambda_{\rm IR} , \Lambda_{\rm UV})  \right]^L .
\ee
Of course, $\lambda^{\rm obs}_{n}$ coincides with the same observable couplings encountered in the previous sections. Here, $\sigma^2 (\Lambda_{\rm IR} , \Lambda_{\rm UV})$ is nothing but the loop integral in Eq.~\eqref{sigma-tot-2-def} {evaluated within the momentum range bounded by physical cutoffs $\Lambda_{\rm IR}$ and $\Lambda_{\rm UV}$, not by comoving cutoffs}. In the particular case of a massless field, one explicitly finds:\footnote{Note that in comoving momentum space the cutoffs are time dependent: $k_{\rm IR}(t) = \Lambda_{\rm IR} a(t)$ and $k_{\rm UV}(t) = \Lambda_{\rm UV} a(t)$. Thus, starting from Eq.~\eqref{sigma-tot-2-def} and integrating the solid angle, one obtains $\sigma^2 (\Lambda_{\rm IR} , \Lambda_{\rm UV})   =\frac{H^2}{4 \pi^2} \int_{\Lambda_{\rm IR} a(t)}^{\Lambda_{\rm UV} a(t)}  \frac{\dd k}{k}  \left( 1 + (k \tau)^2 \right)$. Then, after performing the change of variables $k \to p = k/a(t)$ one obtains the result of Eq.~\eqref{sigma2_phys-phys}.} 
\begin{flalign}
\sigma^2 (\Lambda_{\rm IR} , \Lambda_{\rm UV})  &=\frac{H^2}{4 \pi^2} \int_{\Lambda_{\rm IR}}^{\Lambda_{\rm UV}}  \frac{\dd p}{p}  \left( 1 + \frac{p^2}{H^2} \right)  \notag \\ &= \frac{H^2}{4 \pi^2} \left[  \ln \frac{\Lambda_{\rm UV}}{\Lambda_{\rm IR}}  + \frac{\Lambda_{\rm UV}^2 - \Lambda_{\rm IR}^2}{2 H^2} \right] . \label{sigma2_phys-phys}
\end{flalign}
Now, from Eq.~\eqref{obs-eff}, it is direct to see that
\be \label{V_obs-V_eff}
\V_{\rm obs} (\varphi) = e^{\frac{1}{2} \sigma^2  \frac{\partial^2}{\partial \varphi^2} } \V_{\rm eff} (\varphi) ,
\ee
where $\sigma^2$ stands for $\sigma^2(\Lambda_{\rm IR} , \Lambda_{\rm UV})$. 

The effective potential $\V_{\rm eff} (\varphi)$ must depend on the cutoff scales $\Lambda_{\rm IR}$ and $\Lambda_{\rm UV}$ in order to compensate for the independence of the observable potential $\V_{\rm obs} (\varphi)$ on such arbitrary quantities. Notably, we see that the logarithmic dependence on the ratio $\Lambda_{\rm UV} / \Lambda_{\rm IR}$ coincides with that of the renormalization scale $\mu$ found in the previous section. Given that the renormalization scale $\mu$ is arbitrary, the result of this section can be reconciled with those found in the previous sections through the identification $\frac{\mu}{H} \to  \frac{\Lambda_{\rm UV}}{\Lambda_{\rm IR}} e^{\frac{1}{2 H^2} (\Lambda_{\rm UV}^2 - \Lambda_{\rm IR}^2)}$. Thus, Eq.~\eqref{V_obs-V_eff} determines a flow equation for $\V_{\rm eff} (\varphi)$ in terms of both $\Lambda_{\rm IR}$ and $\Lambda_{\rm UV}$, in the same manner as in Eq.~\eqref{flow}.

In order to be able to perform perturbative computations with this effective theory, $\V_{\rm eff} (\varphi)$ must respect the following perturbative condition:
\be \label{rigid-dS}
|\V_{\rm eff} (\varphi)| \ll H^4 .
\ee
Hence, the flow of $\V_{\rm eff} (\varphi) $ in terms of the cutoff scales could break this perturbative condition for certain values of $\Lambda_{\rm IR}$ and $\Lambda_{\rm UV}$. To appreciate the consequence of this condition, let us again resort to the example of an axionic potential:
\be
\V_{\rm eff} (\varphi) = - A  \cos ( \varphi/f )  .
\ee
The perturbative condition requires $A \ll H^4$. At the same time, thanks to Eq.~\eqref{V_obs-V_eff}, the observable potential can be expressed in terms of $\V_{\rm eff} (\varphi)$ as:
\be
\V_{\rm obs} (\varphi) = - A e^{- \frac{\sigma^2}{2f^2} } \cos (\varphi /f) .
\ee
This implies that the amplitude of the effective potential must depend on the cutoff scales as $A = A_{\rm obs} e^{ \frac{\sigma^2}{2f^2} }$, where $A_{\rm obs}$ is the observable amplitude. Now, the perturbative condition on the effective potential reads $\frac{A_{\rm obs} }{H^4} e^{ \frac{\sigma^2}{2f^2} } \ll 1$, which implies that the theory becomes strongly coupled for values 
\be
\sigma^2 (\Lambda_{\rm IR} , \Lambda_{\rm UV}) \sim 2 f^2 \ln \qty(\frac{H^4}{A_{\rm obs}}) .
\ee
This expression allows us to assess at what infrared or ultraviolet scales the theory becomes non-perturbative. For instance, if we fix $\Lambda_{\rm UV}$ to values not far from $H$, and push $\Lambda_{\rm IR}$ well below $H$, then $\sigma^2\sim H^2 \ln \qty(\frac{H}{\Lambda_{\rm IR}})$ and one learns that the theory becomes non-perturbative at the infrared scale
\be
\Lambda_{\rm IR} \sim  H e^{- 8 \pi \frac{f^2}{H^2} \ln \qty(\frac{H^4}{A_{\rm obs}}) } .
\ee
For momenta below this value of $\Lambda_{\rm IR}$, one cannot trust perturbative computations performed with the effective theory at hand.

\section{Discussion} \label{sec:discussion}
{{
Before concluding, let us comment on some relevant implications of our previous results with a special emphasis on the difference between alternative choices for the infrared cutoff.

\subsection{Physical vs. comoving IR cutoffs} \label{sec:discussion-IR-com}

As we have seen in the previous two sections, the use of physical cutoffs to regularize loops is consistent with dimensional regularization. As mentioned in the introduction, another choice to restrict momenta in loop integrals consists in the use of a physical UV cutoff together with a comoving IR cutoff. Employing this choice in Eq.~\eqref{sigma-tot} for massless modes one finds
\begin{equation} \label{sigma-tot-comov}
\sigma^2(\bar \tau)  =\frac{H^2}{4 \pi^2} \int_{k_0/a(\bar \tau)}^{\Lambda_{\rm UV}}  \frac{\dd p}{p}  \left( 1 +\frac{p^2}{H^2} \right) ,
\end{equation}
where $k_0$ is the infrared cutoff for comoving momenta ($k > k_0$). Upon integration one obtains 
\begin{equation} \label{sigma2_phys-com}
\sigma^2(\bar \tau) = \frac{H^2}{4 \pi^2} \left[  \ln \frac{\Lambda_{\rm UV} a(\bar \tau)}{k_0}  + \frac{\Lambda_{\rm UV}^2 - k_0^2 / a^2(\bar \tau)}{2 H^2} \right] , 
\end{equation}
which clearly differs from Eq.~\eqref{sigma2_phys-phys}. We see that for late times the loop contributes a term proportional to $\ln a(\bar \tau)$. Substituting this expression back into Eq.~\eqref{full-correlator-1} and integrating the $\bar \tau$-vertex variable gives the well known $[\ln a(\tau)]^L$ secular growth implied by loops. 

This way of proceeding (that is, by combining a physical UV cutoff and a comoving IR cutoff) should be analyzed with some care. Note that this choice of cutoffs leads to the appearance of a particular time $\tau_{*}$ where both cutoff scales $\Lambda_{\rm UV} = k_0 / a(\tau_*)$ coincide, signaling the existence of a singular time in a de Sitter background, which a priori doesn't favor any time slice. However, as we have already emphasized, physical observables (in this case correlation functions) must be independent of the choice of cutoffs. That is, the observable correlation function in Eq.~\eqref{full-correlator-2}, with a constant coupling $\lambda_{n}^{\rm obs}$ must be recovered in an effective theory where the ultraviolet cutoff is physical and the infrared cutoff is comoving.

But how is this statement compatible with the conflicting results in Eqs.~\eqref{sigma2_phys-phys} and \eqref{sigma2_phys-com}? The answer is simple: If one works with a comoving cutoff, the Wilsonian coefficients of the effective theory must be such that any time dependence coming from the cutoff is canceled out by the coefficients, making observables independent of the (time-dependent) cutoff. In other words, we could repeat the entire discussion of Sec.~\ref{sec:Wilsonian} with $\Lambda_{\rm IR} = k_0 / a(\bar \tau)$ to find that the resulting correlation functions, being independent of $\Lambda_{\rm IR}$, will not pick up secular growth coming from loops. 

More to the point, working with a comoving cutoff, the effective field theory in Eq.~\eqref{action-2} must now have a time dependent potential $\V_{\rm eff} (\tau , \varphi)$ defined by time dependent Wilsonian coefficients $\lambda_{n}^{\rm eff} (\tau)$. Then, the computation of an $n$-point correlation function in momentum space to first order in $\V_{\rm eff}$ is given by
\begin{widetext}
\be \label{full-correlator-comoving-cutoff}
\langle \varphi_{{\bk}_1} \cdots \, \varphi_{{\bk}_n} \rangle (\tau) = (2 \pi)^3 \delta(\boldsymbol{K}) \frac{1}{H^4}  \times  2\,\text{Im}\Bigg\{ \! \int_{-\infty}^{\tau} \!\! \frac{\dd{\bar\tau}}{{\bar\tau}^{4}}G_+(\bar\tau,k_1)\dotso G_+(\bar\tau,k_n) \sum_{L=0}^\infty \frac{\lambda_{n+2L}^{\rm eff} (\bar \tau)}{L!} \left( \frac{\sigma_{\rm tot}^2(\bar \tau)}{2} \right)^L \! \Bigg\} ,
\ee
\end{widetext}
where $\sigma^2(\bar \tau)$ is the time dependent loop factor already given in Eq.~\eqref{sigma2_phys-com}. Now, it is crucial to appreciate that Eq.~\eqref{full-correlator-comoving-cutoff} cannot differ from Eq.~\eqref{obs-n-pt}; otherwise, the effective theory in Eq.~\eqref{action-2} is invalid. This then implies that 
\be \label{lambda-obs-lambda-eff}
\lambda_n^{\rm obs} = \sum_{L=0}^\infty \frac{\lambda_{n+2L}^{\rm eff} (\bar \tau)}{L!} \left( \frac{\sigma^2(\bar \tau)}{2} \right)^L ,
\ee
where $\lambda_n^{\rm obs}$ is the observable time-independent coupling. This relation tells us that the Wilsonian coefficients must depend on the cutoff scale $k_0/a(\bar \tau)$ (and thus also on time) in such a way that they cancel the dependence of $k_0/a(\bar \tau)$ present in $\sigma_{\rm tot}^2(\bar \tau)$. Our discussions in Sects.~\ref{sec:basic} and~\ref{sec:ren-scheem} precisely show the form of the time dependence of these coefficients or, equivalently, the effective potential. Indeed, Eq.~\eqref{lambda-obs-lambda-eff} implies 
\be \label{V-obs-Veff-t}
\V_{\rm obs} (\varphi) =  e^{  \frac{1}{2} \sigma^2 (\tau) \frac{\partial^2}{\partial \varphi^2}}  \V_{\rm eff} (\tau , \varphi) ,
\ee 
which, together with Eq.~\eqref{weierstrass-obs-b}, leads to the effective potential in terms of the bare potential~\cite{Palma:2023idj}: $\V_{\rm eff} (\varphi, \tau) =  e^{  \frac{1}{2} [ \sigma_{\rm tot}^2 - \sigma^2 (\tau) ]\frac{\partial^2}{\partial \varphi^2}}  \V_{\rm b} (\varphi)$. Inverting Eq.~\eqref{V-obs-Veff-t} and expanding in a Taylor series, one finally finds that the time dependent Wilsonian coefficients expressed in terms of the observable couplings $\lambda^{\rm obs}_n$ are given by:
\be
\lambda_n^{\rm eff} (\bar \tau) = \sum_{L=0}^\infty (-1)^L \frac{\lambda_{n+2L}^{\rm obs}}{L!} \left( \frac{\sigma^2 (\bar \tau) }{2} \right)^L .
\ee
These coefficients guarantee that correlation functions are independent of the cutoff choice.

Readers should be alerted that, in most of the pertinent literature, the computation of correlation functions with comoving cutoffs is done with constant Wilsonian coefficients (that is, independent of $k_0/a(\tau)$). This leads to a spurious time dependence, ultimately inducing the conclusion that loops modify the secular growth of correlators.

\subsection{Correlation functions on superhorizon scales}

Let us take our previous analysis regarding the use of an infrared comoving cutoff one step further. To verify the infrared behavior of the massless scalar field correlation function of Eq.~\eqref{full-correlator-2}, we may evaluate it in the limit of superhorizon external momenta $k_i |\tau| \ll 1$.  
One finds~\cite{Palma:2023idj}:
 \begin{flalign}  \label{correlator-V1-log}
 & \langle \varphi_{{\bk}_1} \cdots \,\varphi_{{\bk}_n} \rangle (\tau)   =  (2 \pi)^3 \delta^{(3)} (\boldsymbol{K}) \notag \\ & \qquad \qquad \times \lambda_n^{\rm obs} \frac{H^{2 (n-2)}}{3 \times 2^{n-1}}  \frac{k_1^3 + \cdots + k_n^3}{k_1^3 \cdots\, k_n^3} \ln \left(K \tau \right) ,
 \end{flalign}
 with $K=\sum k_i$.
This result confirms that correlation functions in comoving momentum space, up to first order in the interaction, have an asymptotic divergence as $\tau \to 0$ proportional to $\ln a(\tau)$. 

To appreciate the importance of the present claims, let us examine the computation of $n$-cumulants\footnote{Here, an $n$-cumulant $\big\langle \varphi^n (\tau) \big\rangle_c$ is the connected part of an $n$-point correlation function in coordinate space evaluated at the coincident limit, where all the coordinates are equal.}  of $\varphi$ on superhorizon scales. As already mentioned in the introduction, observers are limited to a finite range of scales. Consequentially, correlation functions in coordinate space are obtained from correlators in momentum space with the help of window functions filtering the observable range of scales. Consider, for instance, the definition of the so-called infrared field~\cite{Tsamis:2005hd} as
\be
\varphi_L ( {\bx} , \tau) = \int \frac{\dd[3] k}{(2\pi)^3} e^{- i {\bk} \cdot {\bx}} \varphi_{\bk} (\tau) W(k , \tau) ,
\ee
where the window function $W(k , \tau)$, designed to select superhorizon scales $k / a(t) \ll H$, is defined in the following way:
\be \label{window}
W (k,\tau) \equiv \theta(k_*(\tau) - k) \times  \theta(k - k_*(\tau_{\rm i}))  ,
\ee
where $\theta(x)$ is the Heaviside step function and $k_*(\tau) = \alpha H a(\tau)$, for a small dimensionless parameter $\alpha > 0$. Note that $k_*(\tau_{\rm i})$ is a comoving infrared cutoff that limits the number of modes involved in the infrared field in such a way that, at $\tau = \tau_i$, the range of momenta defining $\varphi_L ( {\bx} , \tau)$ is empty. For $\tau > \tau_i$, the range starts to grow as more and more modes enter the definition of $\varphi_L ( {\bx} , \tau)$, implying that the variance $\sigma_L^2 (\tau) \equiv \langle \varphi_L ({\bx}, \tau) \varphi_L ({\by}, \tau) \rangle _{{\bx} = {\by}}$ at zeroth order is given by
\be
\sigma_L^2 (\tau) =\frac{H^2}{4 \pi^2} \int_{\alpha H a(\tau)}^{\alpha H a(\tau_{\rm i})}  \frac{\dd k}{k}  \left( 1 + (k \tau)^2 \right) .
\ee
Integrating this expression and using the fact that $(k \tau)^2 \ll 1$ within the momentum range under consideration, one finds that $\sigma_L^2 (\tau)$} grows proportionally to $\ln \qty( a(\tau) / a(\tau_{\rm i}) )$:
\be
\sigma_L^2 (\tau) = \frac{H^2}{4 \pi^2} \ln  \qty(\frac{a(\tau)}{a(\tau_{\rm i})}) .
\ee

Using Eq.~\eqref{correlator-V1-log}, we can now compute $n$-cumulants to first order in the potential $\mathcal V(\varphi)$. One finds~\cite{Palma:2023uwo}
\be  \label{cumulant-v1}
\Big\langle \varphi_L^n (\tau) \Big\rangle_c  = - \frac{4 \pi^2 n \lambda^{\rm obs}_n }{3 H^4} \sigma_L^{2n} (\tau)  .
\ee
In contrast, if one uses an infrared comoving cutoff, not only to restrict external momenta, but also to cut off momenta appearing in loop integrals involved in the computation of correlation functions, one would obtain time-dependent $n$-cumulants in a way that is sensitive to the number of loops. One finds:
\be  \label{cumulant-v2}
\!\!\!\Big\langle \varphi_L^n (\tau) \Big\rangle_c \!\!= - \frac{4 \pi^2 n}{3H^4} \sigma_L^{2n} (\tau)  \sum_{L=0}^{\infty} \frac{\lambda_{n+2L}}{(n+L) L!} \! \left( \frac{\sigma_{L}^2(\tau)}{2}
 \right)^{L} \!\!\! ,
\ee
where $\lambda_n$ represent ultraviolet (but not infrared) renormalized couplings, and the index $L$, over which the sum is performed, informs us about the number of loops contributing to a given term. 

Because $\big\langle \varphi^n (\tau) \big\rangle_c$ is an observable, it is pertinent to establish the correct procedure to determine the time dependence of $n$-point correlation functions implied by loops. In Ref.~\cite{Palma:2023uwo}, the consequence emerging from the difference between Eqs.~\eqref{cumulant-v1} and \eqref{cumulant-v2} is examined within the context of the stochastic approach to inflation~\cite{Starobinsky:1986fx,Starobinsky:1994bd}. In the context of the present work, let us emphasize that Eq.~\eqref{cumulant-v2} is a standard result obtained in an effective field theory where the infrared cutoff is taken to be comoving, while the ultraviolet cutoff is kept physical, but with constant Wilsonian coefficients. As we just saw in Sec.~\ref{sec:discussion-IR-com}, this way of proceeding leads to spurious time-dependent contributions, which explains the $L$-dependence in Eq.~\eqref{cumulant-v2}. If one carefully considers the time dependence of Wilsonian coefficients, one obtains Eq.~\eqref{full-correlator-2}, which in turn leads to Eq.~\eqref{cumulant-v1}.

\section{Conclusions} \label{sec:conclusions}

There is an ongoing debate on how to regularize infrared divergences arising from loop corrections in de Sitter spacetime. A common approach is to introduce infrared cutoffs to regularize these loop integrals. However, the community remains divided on the appropriate choice for the infrared cutoff, i.e., whether it should be comoving or physical.

To address this issue, we first verified that observable correlation functions, to first order in the scalar field potential $\mathcal{V}(\varphi)$ and to all orders in loops, are indistinguishable from their tree-level counterparts. This result emerges naturally from dimensional regularization and is independent of any specific infrared or ultraviolet cutoffs in momentum space. This is expected, as observable quantities should not depend on arbitrary cutoffs. Consequently, if one chooses to work with cutoffs, the final observable correlation functions must remain independent of that choice. In other words, whether one uses comoving, physical, or alternative cutoffs, a correct calculation will always yield the same observable correlation function.

There are, however, key points to consider. If cutoffs are employed, the effective field theory describing the physics within a bounded energy range will include Wilsonian coefficients that depend on these cutoffs. Crucially, this dependence must cancel out so that observable quantities remain unaffected by the cutoffs. As we have shown, introducing a comoving infrared cutoff inevitably leads to time-dependent Wilsonian coefficients, which are necessary to eliminate the spurious time dependence introduced by the cutoff. For this reason, it is often more reliable to work with physical cutoffs, which avoid the need for time-dependent Wilsonian coefficients. This is perhaps the most significant insight of this article: when computing cumulants (i.e., correlation functions in coordinate space at coincident points), many authors apply the same cutoff to both external and internal momenta. Our results demonstrate that this approach needs special care. A properly computed correlation function in momentum space is independent of the cutoff choice. This in turn indicates that the choice of cutoffs for external momenta is a separate issue that depends on the physical system under consideration. If one insists on using the same cutoffs for both internal and external momenta, one is forced to integrate out infrared modes with a time-dependent (comoving) cutoff, which, in turn, would necessitate the use of time-dependent Wilsonian coefficients, a method that is generally not adopted.

The single-vertex result presented here is suggestive of the more general case with diagrams with two or more vertices: Loops should not modify the leading dependence of diagrams on powers of $\ln a(\tau)$. Of course, the analysis of diagrams with two or more vertices is much more complicated as loops carry external momenta. However, from the present analysis it should be clear that their treatment cannot depend on the nature of cutoffs. If possible, loop integrals should be performed within the dimensional regularization scheme, with ultraviolet and infrared divergences treated in the standard way. A promising avenue to settle the situation with more complicated diagrams is offered by recent developments in integrating general diagrams in de Sitter spacetime~\cite{Xianyu:2022jwk,Qin:2023nhv,Qin:2023bjk}, or large-$N$ techniques~\cite{LopezNacir:2018xto}.

Naturally, the class of theories we are dealing with, being non-renormalizable, will require the introduction of terms (possibly non-local) in order to eliminate infrared divergences without breaking the crucial property of time translation invariance found in the starting theory. Note that here we are in the rigid de Sitter limit, controlled by the perturbative condition in Eq.~\eqref{rigid-dS}, which differs from the case of inflation where time-translation invariance is lost~\cite{Cheung:2007st} as a result of a preferred time set by the slow-rolling inflaton.

\begin{acknowledgments}

We wish to thank Gabriel Mar\'in Mac\^edo for comments. GAP acknowledges support from the Fondecyt Regular project 1210876 (ANID). SS is supported by Thailand NSRF via PMU-B [grant number B37G660013]. JH is supported by ANID-Subdirección del Capital Humano/Magíster Nacional/2023-22231422.

\end{acknowledgments}

\appendix

\section{Feynman rules for correlation functions} \label{app:feynman}

In this appendix, we summarize the Feynman rules allowing one to obtain a correlation function $\big \langle \varphi^n ( {\bk}_1 , \cdots , {\bk}_n ) \big \rangle$ evaluated at a given time $\tau$. We will restrict our attention to a theory of the form in Eq.~\eqref{action-1}. To start with, a single term of the expansion in Eq.~\eqref{eqn:Vb}, proportional to $\lambda_n$, defines two classes of vertices, hereby distinguished by black and white solid dots:
\def\nvertexb{\tikz[baseline=-0.6ex,scale=1.8, every node/.style={scale=1.4}]{
\coordinate (v1) at (0ex,0ex);
\coordinate (phi1) at (-3ex,2ex);
\coordinate (phi2) at (-4ex,0ex);
\coordinate (phi3) at (-3ex,-2ex);
\coordinate (phi4) at (3ex,2ex);
\coordinate (phi5) at (4ex,0ex);
\coordinate (phi6) at (3ex,-2ex);
\draw[thick] (v1) -- (phi1);
\draw[thick] (v1) -- (phi2);
\draw[thick] (v1) -- (phi3);
\draw[thick] (v1) -- (phi4);
\draw[thick] (v1) -- (phi5);
\draw[thick] (v1) -- (phi6);
\filldraw[color=black, fill=black, thick] (v1) circle (0.5ex);
\node[anchor=south] at ($(v1)+(0,-2.5ex)$) {\scriptsize{$\bar\tau$}};
\node[anchor=south] at ($(v1)+(0,+1.0ex)$) {\scriptsize{$\cdots$}};
}
}
\def\nvertexw{\tikz[baseline=-0.6ex,scale=1.8, every node/.style={scale=1.4}]{
\coordinate (v1) at (0ex,0ex);
\coordinate (phi1) at (-3ex,2ex);
\coordinate (phi2) at (-4ex,0ex);
\coordinate (phi3) at (-3ex,-2ex);
\coordinate (phi4) at (3ex,2ex);
\coordinate (phi5) at (4ex,0ex);
\coordinate (phi6) at (3ex,-2ex);
\draw[thick] (v1) -- (phi1);
\draw[thick] (v1) -- (phi2);
\draw[thick] (v1) -- (phi3);
\draw[thick] (v1) -- (phi4);
\draw[thick] (v1) -- (phi5);
\draw[thick] (v1) -- (phi6);
\filldraw[color=black, fill=white, thick] (v1) circle (0.5ex);
\node[anchor=south] at ($(v1)+(0,-2.5ex)$) {\scriptsize{$\bar\tau$}};
\node[anchor=south] at ($(v1)+(0,+1.0ex)$) {\scriptsize{$\cdots$}};
}
}
\begin{flalign}
    & \nvertexb \;\;\;\;\;\;  \longrightarrow  \notag \\ & \;\;\;\;\;\;  - i \lambda_n \int^{\tau}_{- \infty} \! \de\bar\tau \, a^4(\bar\tau) (2 \pi)^3 \delta^{(3)} \qty(\sum_i k_{i}) \bigg[ \cdots \bigg] , \\
    & \nvertexw \;\;\;\;\;\;  \longrightarrow  \notag \\ &  \;\;\;\;\;\; + i \lambda_n \int^{\tau}_{- \infty} \! \de\bar\tau \, a^4(\bar\tau) (2 \pi)^3 \delta^{(3)} \qty(\sum_i k_{i}) \bigg[ \cdots \bigg] ,
\end{flalign}
where $\delta^{(3)} (\sum_i k_{i})$ denotes a Dirac delta function enforcing conservation of momenta flowing into the vertex through the legs. Each vertex is characterized by a time label $\bar\tau$ that must be integrated from $-\infty$ up until the time $\tau$ at which $n$-point correlation functions are evaluated. The square brackets on the right hand side of the previous equations indicate that any function of $\bar\tau$ must be integrated in this way. 

Given that we have two classes of vertices, the theory will contain four types of internal propagators. The corresponding internal propagators are given by
\def\propbb{\tikz[baseline=-0.6ex,scale=1.8, every node/.style={scale=1.4}]{
\coordinate (tau1) at (-4ex,0ex);
\coordinate (tau2) at (4ex,0ex);
\draw[thick] (tau1) -- (tau2);
\filldraw[color=black, fill=black, thick] (tau1) circle (0.5ex);
\node[anchor=south] at ($(tau1)+(0,0.5ex)$) {\scriptsize{$\tau_1$}};
\filldraw[color=black, fill=black, thick] (tau2) circle (0.5ex);
\node[anchor=south] at ($(tau2)+(0,0.5ex)$) {\scriptsize{$\tau_2$}};
}
}
\def\propww{\tikz[baseline=-0.6ex,scale=1.8, every node/.style={scale=1.4}]{
\coordinate (tau1) at (-4ex,0ex);
\coordinate (tau2) at (4ex,0ex);
\draw[thick] (tau1) -- (tau2);
\filldraw[color=black, fill=white, thick] (tau1) circle (0.5ex);
\node[anchor=south] at ($(tau1)+(0,0.5ex)$) {\scriptsize{$\tau_1$}};
\filldraw[color=black, fill=white, thick] (tau2) circle (0.5ex);
\node[anchor=south] at ($(tau2)+(0,0.5ex)$) {\scriptsize{$\tau_2$}};
}
}
\def\propbw{\tikz[baseline=-0.6ex,scale=1.8, every node/.style={scale=1.4}]{
\coordinate (tau1) at (-4ex,0ex);
\coordinate (tau2) at (4ex,0ex);
\draw[thick] (tau1) -- (tau2);
\filldraw[color=black, fill=black, thick] (tau1) circle (0.5ex);
\node[anchor=south] at ($(tau1)+(0,0.5ex)$) {\scriptsize{$\tau_1$}};
\filldraw[color=black, fill=white, thick] (tau2) circle (0.5ex);
\node[anchor=south] at ($(tau2)+(0,0.5ex)$) {\scriptsize{$\tau_2$}};
}
}
\def\propwb{\tikz[baseline=-0.6ex,scale=1.8, every node/.style={scale=1.4}]{
\coordinate (tau1) at (-4ex,0ex);
\coordinate (tau2) at (4ex,0ex);
\draw[thick] (tau1) -- (tau2);
\filldraw[color=black, fill=white, thick] (tau1) circle (0.5ex);
\node[anchor=south] at ($(tau1)+(0,0.5ex)$) {\scriptsize{$\tau_1$}};
\filldraw[color=black, fill=black, thick] (tau2) circle (0.5ex);
\node[anchor=south] at ($(tau2)+(0,0.5ex)$) {\scriptsize{$\tau_2$}};
}
}
\begin{flalign}
\propbb  \;\;\;\;\;\;  &\longrightarrow  \;\;\;\;\;\; G_{++} ( \tau_1 , \tau_2,k) , \\
\propww  \;\;\;\;\;\; &\longrightarrow  \;\;\;\;\;\; G_{--} (\tau_1 , \tau_2,k) , \\
\propbw  \;\;\;\;\;\; &\longrightarrow \;\;\;\;\;\;  G_{+-} (\tau_1 , \tau_2,k) ,\\
\propwb  \;\;\;\;\;\; &\longrightarrow \;\;\;\;\;\;  G_{-+} (\tau_1 , \tau_2,k) .
\end{flalign}
The analytical expressions for the quantities appearing at the right hand side of the previous assignments are given by:
\bea
G_{++} (\tau_1 , \tau_2,k)  &=& f(k , \tau_1) f^*(k , \tau_2) \theta (\tau_1 - \tau_2) \notag \\ & &\, +  f^*(k , \tau_1) f(k , \tau_2) \theta (\tau_2 - \tau_1) , \\
G_{--} (\tau_1 , \tau_2,k)  &=& f^*(k , \tau_1) f(k , \tau_2) \theta (\tau_1 - \tau_2) \notag \\ & &\, +  f(k , \tau_1) f^*(k , \tau_2) \theta (\tau_2 - \tau_1) , \\
G_{+-} (\tau_1 , \tau_2,k)  &=& f^*(k , \tau_1) f(k , \tau_2) , \\
G_{-+} (\tau_1 , \tau_2,k)  &=& f(k , \tau_1) f^*(k , \tau_2) .
\eea
Just as in the case of vertices, it can be verified explicitly that propagators with given colors (black or white) at their ends are the complex conjugates of propagators with the opposite colors.

Furthermore, the vertices (evaluated at times $\tau_1$, $\tau_2$, $\tau_3$, etc.) must be connected to a surface labeled with the time $\tau$ through bulk-to-boundary propagators. These receive the following assignments:
\def\propbf{\tikz[baseline=-0.6ex,scale=1.8, every node/.style={scale=1.4}]{
\coordinate (tau) at (-4ex,0ex);
\coordinate (phi) at (4ex,0ex);
\draw[thick] (tau) -- (phi);
\filldraw[color=black, fill=black, thick] (tau) circle (0.5ex);
\node[anchor=south] at ($(tau)+(0,0.5ex)$) {\scriptsize{$\bar{\tau}$}};
\pgfmathsetmacro{\arista}{0.08}
\filldraw[color=black, fill=white, thick] ($(phi)-(\arista,\arista)$) rectangle ($(phi)+(\arista,\arista)$);
\node[anchor=south] at ($(phi)+(0,0.5ex)$){\scriptsize{$\tau$}};
}
}
\def\propwf{\tikz[baseline=-0.6ex,scale=1.8, every node/.style={scale=1.4}]{
\coordinate (tau) at (-4ex,0ex);
\coordinate (phi) at (4ex,0ex);
\draw[thick] (tau) -- (phi);
\filldraw[color=black, fill=white, thick] (tau) circle (0.5ex);
\node[anchor=south] at ($(tau)+(0,0.5ex)$) {\scriptsize{$\bar\tau$}};
\pgfmathsetmacro{\arista}{0.08}
\filldraw[color=black, fill=white, thick] ($(phi)-(\arista,\arista)$) rectangle ($(phi)+(\arista,\arista)$);
\node[anchor=south] at ($(phi)+(0,0.5ex)$){\scriptsize{$\tau$}};
}
}
\begin{flalign}
  \propbf   \;\;\;\;\;\;  &\longrightarrow  \;\;\;\;\;\;  G_{+}  (\bar\tau,k) , \\
  \propwf  \;\;\;\;\;\;  &\longrightarrow  \;\;\;\;\;\;  G_{-}  (\bar\tau,k) .
\end{flalign}
Note that it is unnecessary to assign colors to the square defining the end point evaluated at $\tau$. The analytical expressions for these bulk-to-boundary propagators read:
\bea
G_{+}  (\bar\tau,k) &=& f^*(k , \bar\tau) f(k , \tau) ,  \\
G_{-}  (\bar\tau,k) &=& f(k , \bar\tau) f^*(k , \tau)  .
\eea
Every internal momentum flowing through propagators must be integrated with $\int \frac{\dd[3] k}{(2 \pi)^3}$. Then, the correlation function $\big \langle \varphi^n ( {\bk}_1 , \cdots , {\bk}_n ) \big \rangle$ corresponds to the summation of every diagram with $n$ external legs, truncated to the desired order.

To finish, let us consider a general 1-vertex diagram with $n$ external lines and $L$ loops. This should be written as the sum of two diagrams with black and white vertices respectively:
\def\Oconnectedarbitraryloopsblack{\tikz[baseline=-1.4ex]{
\coordinate (P) at (0,-3ex);
\coordinate (C) at (-9ex,4ex);
\draw (C) circle (0.01ex) node[anchor=east]{\footnotesize{$\tau$}};
\draw[thick,dashed] (-9ex,4ex) -- (9ex,4ex);
\draw[thick] (0,-3ex) -- (-8ex,4ex);
\draw[thick] (0,-3ex) -- (-4ex,4ex);
\draw[thick] (0,-3ex) -- (4ex,4ex);
\draw[thick] (0,-3ex) -- (8ex,4ex);
\filldraw[color=black, fill=black, thick] (-1.2ex,2ex) circle (0.1ex);
\filldraw[color=black, fill=black, thick] (0ex,2ex) circle (0.1ex);
\filldraw[color=black, fill=black, thick] (1.2ex,2ex) circle (0.1ex);
\node at ($(P) + (-0.8ex,0)$) [anchor=east]{\footnotesize{$\lambda^{\text{b}}_{n+2L},\,\bar\tau$}};
\filldraw[color=black, fill=white, thick] (-8.5ex,3.5ex) rectangle (-7.5ex,4.5ex) node[anchor=south]{\footnotesize{$k_1$}};
\filldraw[color=black, fill=white, thick] (-4.5ex,3.5ex) rectangle (-3.5ex,4.5ex) node[anchor=south]{\footnotesize{$k_2$}};
\filldraw[color=black, fill=white, thick] (3.5ex,3.5ex) rectangle (4.5ex,4.5ex) node[anchor=south]{\footnotesize{$k_{n-1}$}};
\filldraw[color=black, fill=white, thick] (7.5ex,3.5ex) rectangle (8.5ex,4.5ex) node[anchor=south]{\footnotesize{$k_n$}};
\draw[thick,scale=3] (0,-1ex)  to[in=-100,out=-140,loop] (0,-1ex);
\draw[thick,scale=3] (0,-1ex)  to[in=-50,out=-90,loop] (0,-1ex);
\draw[thick,scale=3] (0,-1ex)  to[in=30,out=-10,loop] (0,-1ex);
\pgfmathsetmacro{\distanceone}{0.8}
\pgfmathsetmacro{\angleone}{315}
\pgfmathsetmacro{\distancetwo}{0.8}
\pgfmathsetmacro{\angletwo}{330}
\pgfmathsetmacro{\distancethree}{0.8}
\pgfmathsetmacro{\anglethree}{345}
\coordinate (Qone) at ($(P) + (\angleone:\distanceone)$);
\filldraw[color=black, fill=black, thick] (Qone) circle (0.1ex);
\coordinate (Qtwo) at ($(P) + (\angletwo:\distancetwo)$);
\filldraw[color=black, fill=black, thick] (Qtwo) circle (0.1ex) node[anchor=west]{\scriptsize{$L$ loops}};
\coordinate (Qthree) at ($(P) + (\anglethree:\distancethree)$);
\filldraw[color=black, fill=black, thick] (Qthree) circle (0.1ex);
\filldraw[color=black, fill=black] (P) circle (0.8ex);
}
}
\def\Oconnectedarbitraryloopswhite{\tikz[baseline=-1.4ex]{
\coordinate (P) at (0,-3ex);
\coordinate (C) at (-9ex,4ex);
\draw (C) circle (0.01ex) node[anchor=east]{\footnotesize{$\tau$}};
\draw[thick,dashed] (-9ex,4ex) -- (9ex,4ex);
\draw[thick] (0,-3ex) -- (-8ex,4ex);
\draw[thick] (0,-3ex) -- (-4ex,4ex);
\draw[thick] (0,-3ex) -- (4ex,4ex);
\draw[thick] (0,-3ex) -- (8ex,4ex);
\filldraw[color=black, fill=black, thick] (-1.2ex,2ex) circle (0.1ex);
\filldraw[color=black, fill=black, thick] (0ex,2ex) circle (0.1ex);
\filldraw[color=black, fill=black, thick] (1.2ex,2ex) circle (0.1ex);
\node at ($(P) + (-0.8ex,0)$) [anchor=east]{\footnotesize{$\lambda^{\text{b}}_{n+2L},\,\bar\tau$}};
\filldraw[color=black, fill=white, thick] (-8.5ex,3.5ex) rectangle (-7.5ex,4.5ex) node[anchor=south]{\footnotesize{$k_1$}};
\filldraw[color=black, fill=white, thick] (-4.5ex,3.5ex) rectangle (-3.5ex,4.5ex) node[anchor=south]{\footnotesize{$k_2$}};
\filldraw[color=black, fill=white, thick] (3.5ex,3.5ex) rectangle (4.5ex,4.5ex) node[anchor=south]{\footnotesize{$k_{n-1}$}};
\filldraw[color=black, fill=white, thick] (7.5ex,3.5ex) rectangle (8.5ex,4.5ex) node[anchor=south]{\footnotesize{$k_n$}};
\draw[thick,scale=3] (0,-1ex)  to[in=-100,out=-140,loop] (0,-1ex);
\draw[thick,scale=3] (0,-1ex)  to[in=-50,out=-90,loop] (0,-1ex);
\draw[thick,scale=3] (0,-1ex)  to[in=30,out=-10,loop] (0,-1ex);
\pgfmathsetmacro{\distanceone}{0.8}
\pgfmathsetmacro{\angleone}{315}
\pgfmathsetmacro{\distancetwo}{0.8}
\pgfmathsetmacro{\angletwo}{330}
\pgfmathsetmacro{\distancethree}{0.8}
\pgfmathsetmacro{\anglethree}{345}
\coordinate (Qone) at ($(P) + (\angleone:\distanceone)$);
\filldraw[color=black, fill=black, thick] (Qone) circle (0.1ex);
\coordinate (Qtwo) at ($(P) + (\angletwo:\distancetwo)$);
\filldraw[color=black, fill=black, thick] (Qtwo) circle (0.1ex)  node[anchor=west]{\scriptsize{$L$ loops}};
\coordinate (Qthree) at ($(P) + (\anglethree:\distancethree)$);
\filldraw[color=black, fill=black, thick] (Qthree) circle (0.1ex);
\filldraw[color=black, thick, fill=white] (P) circle (0.8ex);
}
}
\begin{widetext}
\be
\Oconnectedloopstwo = \Oconnectedarbitraryloopsblack  + \Oconnectedarbitraryloopswhite .
\ee
\end{widetext}
Following the rules just outlined, and taking into account the appropriate symmetry factors, one then obtains the expression presented in Eq.~\eqref{eq:arbitrary-diagram}.

\section{Proof of Equation \eqref{solution-div-1}} \label{app:proof}

Here, we show how to obtain the solution in Eq.~\eqref{solution-div-1} to the system of algebraic equations in Eq.~\eqref{algebraic-system} via induction. Written more explicitly, this system has the form
\begin{widetext}
\begin{eqnarray} 
 \lambda^{(1)}_{n} + \left( \frac{\sigma_{\rm tot}^2}{2} \right)  \lambda^{(0)}_{n+2}   &=& 0 ,  \label{general-1}
 \\
\lambda^{(2)}_{n} +  \left( \frac{\sigma_{\rm tot}^2}{2} \right)  \lambda^{(1)}_{n+2} +  \frac{1}{2!} \left( \frac{\sigma_{\rm tot}^2}{2} \right)^{2}  \lambda^{(0)}_{n+4} &=& 0 ,
\label{general-2}
\\
\lambda^{(3)}_{n} +  \left( \frac{\sigma_{\rm tot}^2}{2} \right)  \lambda^{(2)}_{n+2} +  \frac{1}{2!} \left( \frac{\sigma_{\rm tot}^2}{2} \right)^{2}  \lambda^{(1)}_{n+4} +  \frac{1}{3!} \left( \frac{\sigma_{\rm tot}^2}{2} \right)^{3}  \lambda^{(0)}_{n+6}&=& 0 ,
\label{general-3}
\\
\vdots \notag
\\
\lambda^{(L)}_{n} +  \left( \frac{\sigma_{\rm tot}^2}{2} \right)  \lambda^{(L-1)}_{n+2} +  \frac{1}{2!} \left( \frac{\sigma_{\rm tot}^2}{2} \right)^{2}  \lambda^{(L-2)}_{n+4} + \cdots +  \frac{1}{L!} \left( \frac{\sigma_{\rm tot}^2}{2} \right)^{L}  \lambda^{(0)}_{n+2L}&=& 0, \label{general-L}
\end{eqnarray}
\end{widetext}
and so on and so forth. The first equation (Eq.~\eqref{general-1}) has the solution $ \lambda^{(1)}_{n} = - \left( \frac{\sigma_{\rm tot}^2}{2} \right)  \lambda^{(0)}_{n+2} $. This solution, in turn, implies that $ \lambda^{(1)}_{n+2} = - \left( \frac{\sigma_{\rm tot}^2}{2} \right)  \lambda^{(0)}_{n+4} $, from where we can solve the second equation---Eq.~\eqref{general-2}---to obtain $\lambda_{n}^{(2)} = \frac{1}{2} \left( \frac{\sigma_{\rm tot}^2}{2} \right)^2 \lambda^{(0)}_{n+4}$. This logic can be carried out to an arbitrary degree with induction. Thus, let us assume that $\lambda_n^{(p)} = (-1)^p \frac{1}{p!} \left( \frac{\sigma_{\rm tot}^2}{2} \right)^p \lambda_{n+2p}^{(0)}$ for all $p \leq L-1$ and show that it also holds true for $p = L$. To proceed, notice that $\lambda_n^{(L)}$ satisfies Eq.~\eqref{general-L}, where every $\lambda_n^{(p)}$ with $p \leq L-1$ is known. Then, replacing $\lambda^{(p)}_{n + 2 (L-p)} = (-1)^{p} \frac{1}{p!} \left( \frac{\sigma_{\rm tot}^2}{2} \right)^p \lambda^{(0)}_{n+ 2 L}$ for $p \leq L-1$, Eq.~\eqref{general-L} gives us back:
\be
\lambda_{n}^{(L)} = - \left( \sum_{p=0}^{L-1} (-1)^p \frac{1}{(L - p)! p!}  \right)  \left( \frac{\sigma_{\rm tot}^2}{2} \right)^{L} \lambda^{(0)}_{n+2L} .
\ee
Thanks to Pascal's formula for binomial coefficients, it is possible to show that:
\be
\sum_{p=0}^{L-1} (-1)^p \frac{1}{(L - p)! p!}  = (-1)^{L-1} \frac{1}{L!} .
\ee
This finally leads to the result in Eq.~\eqref{solution-div-1}:
\be
\lambda_{n}^{(L)} = (-1)^{L} \frac{1}{L!} \left( \frac{\sigma_{\rm tot}^2}{2} \right)^{L} \lambda^{(0)}_{n+2L} .
\ee

\bibliography{bibliography}

\end{document}